\documentclass[onecolumn,preprint]{emulateapj}

\usepackage{mathtext,bbm,amsmath,amsfonts,amssymb,indentfirst,syntonly,graphicx}
\usepackage[english]{babel}
\usepackage{slashbox}
\usepackage{calc}
\usepackage{tikz}
\usepackage{latexsym,graphicx,bbm}
\usepackage{epstopdf}

\newcommand{\beq}{\begin{eqnarray}}
\newcommand{\eeq}{\end{eqnarray}}

\begin{document}

\title{Model-independently calibrating the luminosity correlations of gamma-ray bursts using deep learning}
\author{Li Tang\altaffilmark{1,2}, Xin Li\altaffilmark{2}, Hai-Nan Lin\altaffilmark{2,*}, Liang Liu\altaffilmark{1,2}}
\affil{\altaffilmark{1}Department of Math and Physics, Mianyang Normal University, Mianyang 621000, China\\
\altaffilmark{2}Department of Physics, Chongqing University, Chongqing 401331, China\\}
\altaffiltext{*}{corresponding author: linhn@cqu.edu.cn}

\begin{abstract}
  Gamma-ray bursts (GRBs) detected at high redshift can be used to trace the Hubble diagram of the Universe. However, the distance calibration of GRBs is not as easily as that of type Ia supernovae (SNe Ia). For the calibrating method based on the empirical luminosity correlations, there is an underlying assumption that the correlations should be universal over the whole redshift range. In this paper, we investigate the possible redshift dependence of six luminosity correlations with a completely model-independent deep learning method. We construct a network combining the Recurrent Neural Networks (RNN) and the Bayesian Neural Networks (BNN), where RNN is used to reconstruct the distance-redshift relation by training the network with the Pantheon compilation, and BNN is used to calculate the uncertainty of the reconstruction. Using the reconstructed distance-redshift relation of Pantheon, we test the redshift dependence of six luminosity correlations by dividing the full GRB sample into two subsamples (low-$z$ and high-$z$ subsamples), and find that only the $E_p-E_{\gamma}$ relation has no evidence for redshift dependence. We use the $E_p-E_{\gamma}$ relation to calibrate GRBs, and the calibrated GRBs give tight constraint on the flat $\Lambda$CDM model, with the best-fitting parameter $\Omega_{\rm M}$=0.307$^{+0.065}_{-0.073}$.
\end{abstract}

\section{Introduction}\label{intro}

The accelerating expansion of the universe is first found from the fact that the luminosity of type Ia supernovae (SNe Ia) is dimmer than expected \citep{Riess:1998,Perlmutter:1999}. Since then, many SNe Ia datasets have been compiled to cosmological researches \citep{Suzuki:2012,Betoule:2014,Scolnic:2018}. However, due to the limited luminosity, most observable SNe Ia are at redshift $z\lesssim 2$. With these low redshift data, cosmological models cannot be unambiguously distinguished \citep{Zhu:2008,Dolgov:2014,Tutusaus:2016,Wei:2016}. Among these SNe Ia samples, the most up-to-date Pantheon compilation \citep{Scolnic:2018} is the largest sample and the redshift of the furthest SNe reaches up to $z\sim$ 2.3, but the number of SNe whose redshift is larger than 1.4 is only six. Actually, the subtle difference between cosmological models in low redshift range would be remarkable in high redshift range. For various cosmological models, such as $\Lambda$CDM model, wCDM model, Chevallier-Polarsky-Linder (CPL) model, holographic dark energy (HDE) model and generalised chaplygin gas (GCG) model, the evolutions of dark energy equation of state are discrepant in high redshift \citep{Chevallier:2001,Wang:2017,Pan:2020,Escamilla-Rivera et al.:2020}. Thus, it is important to extend the Hubble diagram to high redshift range. Gamma-ray bursts (GRBs), as the most energetic explosions in the universe, is detectable up to very high redshift \citep{Cucchiara:2011}. Therefore, it is possible to use GRBs as standard candles to trace the Hubble diagram at high redshift. Combining GRBs with other standard candles, the cosmological parameters can be tightly constrained \citep{Friedman:2005,Wang:2006,Izzo:2009,Liang:2011,Bloom:2003,Xu:2005,Wei:2009,Demianski:2011,Cai:2013,Chang:2014,Lin:2016}. However, it is not easy to calibrate the distance of GRBs due to the lack of knowledge on the explosion mechanism.

Based on the correlations between various observables of the prompt or afterglow emission, several methods have been proposed to calibrate the distance of GRBs \citep{Dai:2004,Ghirlanda:2004,Firmani:2005,Schaefer:2007,Liang:2008,Wei:2010,Liu:2015}. Most calibrating methods rely on one of the following 2-dimensional empirical luminosity calibrations found in long GRBs:
the correlation between spectrum lag and isotropic peak luminosity ($\tau_{\rm lag} - L$ relation) \citep{Norris:2000},
the correlation between time variability and isotropic peak luminosity ($V - L$ relation) \citep{Fenimore:2000},
the correlation between peak energy and isotropic peak luminosity ($E_p - L$ relation) \citep{Yonetoku:2004},
the correlation between peak energy and collimation-corrected energy ($E_p - E_{\gamma}$ relation) \citep{Ghirlanda:2004},
the correlation between minimum rise time of light curve and isotropic peak luminosity ($\tau_{\rm RT}- L$ relation) \citep{Schaefer:2007} and
the correlation between the peak energy of $\nu F_{\nu}$ spectrum and isotropic equivalent energy ($E_p - E_{\rm iso}$ relation) \citep{Amati:2002}.

What is noteworthy is that, all of the above correlations depend on a certain cosmological model, thus leading to the circularity problem when use the calibrated GRBs to constrain cosmological models. Several model-independent methods have been proposed by using distance ladders to calibrate GRBs \citep{Liang:2008,Wei:2009,Wei:2010,Liu:2015}. Using cubic interpolation or other approximations from SNe Ia dataset, one can first calculates the distance for low-$z$ ($z<$ 1.4) GRBs to derive the empirical luminosity correlations. Then by extrapolating the correlations to high-$z$ ($z>$ 1.4) GRBs, one can inversely derive the distance for high-$z$ GRBs. However, in addition to the dependency of a certain approximation form, there is an underlying assumption in this method that the luminosity correlations are universal over all redshifts. Many works have been devoted to test the possible redshift dependence of luminosity correlations \citep{Li:2007,Basilakos:2008,Wang:2011,Lin:2015,Lin:2016}. Assuming a flat $\Lambda$CDM model with different cosmological parameters, \cite{Basilakos:2008} and \cite{Wang:2011} investigated the above six empirical luminosity correlations in four redshift bins, and found no statistically significant evidence for redshift evolution. While \cite{Lin:2016} rechecked these correlation in two redshift bins, and found moderate evidence ($>3\sigma$) for the redshift evolution in four out of six correlations.

All the above works to check the redshift dependence of GRB luminosity correlations still depend on cosmological model. In this paper, we will investigate the redshift dependence of these correlations with a model-independent method, i.e. the deep learning, which is one of the most exciting areas in machine learning. Deep learning is tackling large and highly complex machine learning tasks by training deep neural networks constructed with an input layer to receive the features, several hidden layers to transform the information from the previous layer, and an output layer to export the target, where each layer contains hundreds of nonlinear processing neurons \citep{Aurelien:2017}. Recently, deep learning method has been widely employed in cosmological researches, such as predicting galaxy morphology \citep{Dieleman:2015}, learning the universe at scale \citep{Mathuriya:2018}, constraining cosmological dark energy \citep{Escamilla-Rivera et al.:2020}, and so on. In this paper, we proposed a method to calibrate the luminosity correlations of GRBs using deep learning method, which is completely independent on the cosmological model. We first combine the Recurrent Neural Networks (RNN) and the Bayesian Neural Networks (BNN) to reconstruct the distance-redshift relation from the Pantheon sample up to the highest redshift of GRB dataset, then we test the redshift dependence of GRB luminosity correlations by dividing the full GRB sample into low-$z$ ($z\leq1.4$) and high-$z$ ($z>1.4$) subsamples. Comparing with previous works \citep{Basilakos:2008,Wang:2011,Lin:2016}, our method is model-independent and only relies on the training data, i.e. the Pantheon compilation. Without any assumption about the cosmological model or about the specific form of distance-redshift relation of SNe Ia, we directly calibrate the distance of GRBs from the Pantheon sample with deep learning.

The rest of the paper is organized as follows: In Section \ref{sec:network}, we introduce the architecture of the RNN+BNN network. In Section \ref{sec:Reconstruction}, we use the network to reconstruct the distance-redshift relation from Pantheon supernovae data set. In Section \ref{sec:Test}, we use the reconstructed distance-redshift relation to test the possible redshift dependence of luminosity correlations of GRBs. Finally, discussions and conclusion are given in Section \ref{sec:conclusion}.

\section{The architecture of the neural network}\label{sec:network}

Before testing the redshift dependence of luminosity correlations of GRBs, we firstly introduce the process of reconstructing the luminosity distance of SNe with deep learning. Based on the training Artificial Neural Networks (ANN), such as Convolutional Neural Networks (CNN), Recurrent Neural Networks (RNN) and Bayesian Neural Networks (BNN), deep learning is versatile, powerful and scalable in tackling complex problems such as classifying billions of images, recognizing speech, detecting subtle patterns in data, etc \citep{Aurelien:2017}. Recently, the application of deep learning in cosmological research is very extensive and successful \citep{Dieleman:2015,Mathuriya:2018,Escamilla-Rivera et al.:2020}. Following the work of \cite{Escamilla-Rivera et al.:2020}, we reconstruct the distance moduli from the Pantheon compilation \citep{Scolnic:2018} with RNN+BNN. In this process, the reconstruction of distance only depends on the Pantheon dataset, and without any assumption on the cosmological model.

RNN is a class of nets which can predict the future from the complex sequential information without any model assumption, but is incapable of estimating the uncertainty of target. This shortcoming can be fixed up with BNN. Therefore, our neural network is composed of RNN and BNN, the details of which are described bellow.

\begin{figure}[htbp]
  \centering
  \includegraphics[width=0.8\textwidth]{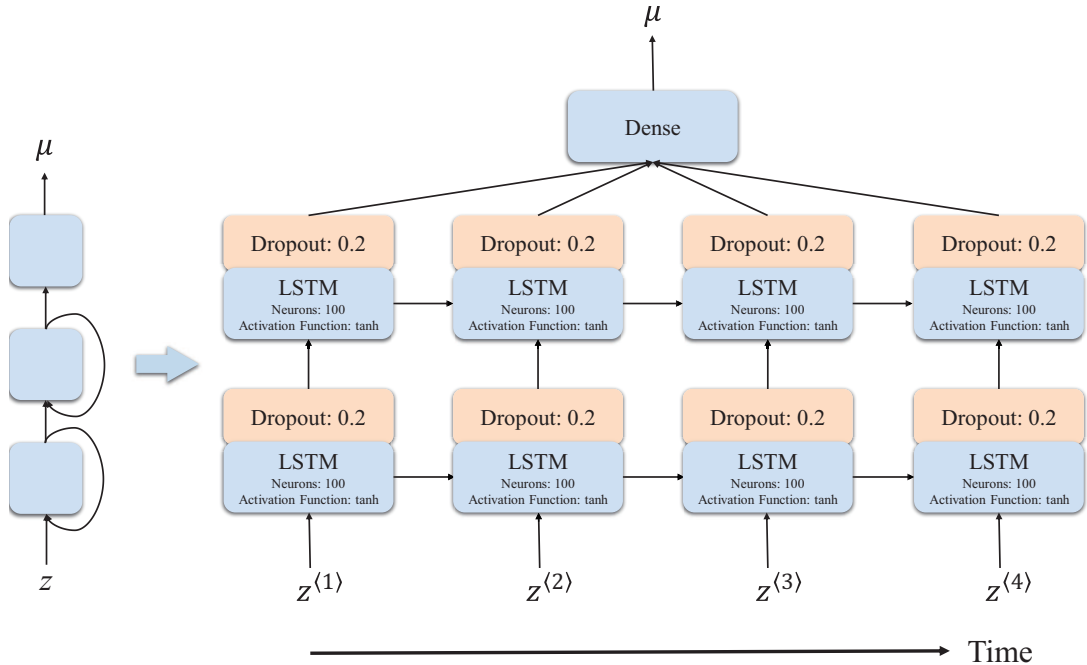}
  \caption{\small{The architecture of our network with one hidden layer (left), unrolled through time step $t$=4 (right). In the unrolled network, each column is one of the $t$ time steps, while the three rows from bottom to top represent input layer, hidden layer and output layer, respectively. The first two layers with tanh activation function consist of LSTM cell containing 100 neurons, while the output layer is a fully-connected (dense) layer. To avoid overfitting, the dropout technique is employed between LSTM and its next layers, and we set the dropout rate to 0.2.}}\label{fig_ourRNN}
\end{figure}

RNN is one of the supervised learning algorithm by training the neural networks with the real data, reaching an ideal network characterizing the relationship between the target and the feature by minimizing the loss function \citep{Aurelien:2017}. In our work, the Pantheon data set is used as the training data, in which the redshift is the feature and the distance module is the target. In RNN, the activation not only flows from the input layer to the output layer, but also has connections pointing backward. The architecture of RNN is shown in Figure \ref{fig_ourRNN}. In the unrolled RNN, the neurons at each time step $t$ receive the inputs as well as the outputs from the previous time step \citep{Aurelien:2017}. In the neural network, the loss function is used to depict the difference between the targets and the predicts. We adopt the Mean Squared Error (MSE) function as the loss function and find the minimum with the Adam optimizer.

Handling long sequences, the training of RNN will take a long time and the information of initial inputs will gradually fades away \citep{Aurelien:2017}. Thus, we adopt the time step $t=4$ to alleviate the long training time, and use the most popular basic cell called Long Short-Term Memory (LSTM) cell to solve the problem of information loss. RNN with LSTM cell is aware of what to store, throw away and read. The computations of LSTM are
\begin{equation}
i^{<t>}=\sigma\left(W_{xi}^T \cdot x^{<t>}+W_{hi}^T \cdot h^{<t-1>}+b_i\right)
\end{equation}
\begin{equation}
f^{<t>}=\sigma\left(W_{xf}^T \cdot x^{<t>}+W_{hf}^T \cdot h^{<t-1>}+b_f\right),
\end{equation}
\begin{equation}
o^{<t>}=\sigma\left(W_{xo}^T \cdot x^{<t>}+W_{ho}^T \cdot h^{<t-1>}+b_o\right),
\end{equation}
\begin{equation}
g^{<t>}=A_f\left(W_{xg}^T \cdot x^{<t>}+W_{hg}^T \cdot h^{<t-1>}+b_g\right),
\end{equation}
\begin{equation}
c^{<t>}=f^{<t>}\otimes c^{<t-1>}+i^{<t>}\otimes g^{<t>},
\end{equation}
\begin{equation}
y^{<t>}=h^{<t>}=o^{<t>}\otimes A_f\left(c^{<t>}\right),
\end{equation}
where $\sigma$ is the sigmoid function that outputs a value between 0 and 1, $t$ is the time step referring to the current sequence (for example $t=1$ for the first redshift). The superscript $<t>$ indicates a vector of steps $t$, the superscript $T$ is the transpose of the matrix, the dot is matrix product and $\otimes$ is direct product. $x^{<t>}$ and $y^{<t>}$ are respectively the current input and output vectors. $h^{<t>}$ and $c^{<t>}$ are respectively the short-term state and long-term state of LSTM cells. $A_f$ is an activation function to make the network be capable of solving complex tasks by introducing the non-linearity to network. In our work, we use the tanh activation function, which is defined as
\begin{equation}
A_{f_{\rm Tanh}}={\rm tanh}(x)=\frac{e^x-e^{-x}}{e^x+e^{-x}}.
\end{equation}
There are four connected layers playing different roles, where the main layer that outputs $g^{<t>}$ analyzes the current inputs $x^{<t>}$ and the previous state $h^{<t-1>}$, the rest three layers are $gate \ controllers$: (a) $Input \ gate$ controlled by $i^{<t>}$ determines which parts of $g^{<t>}$ should be added to $c^{<t>}$, (b) $Forget \ gate$ controlled by $f^{<t>}$ determines which parts of $c^{<t>}$ should be abandoned, (c) $Output \ gate$ controlled by $o^{<t>}$ determines which parts of $c^{<t>}$ should be output. It can be easily found that, these gate controllers are related to the logistic activation function $\sigma$, thus they would close the gate if output 0 and open it if output 1.
$W_{xi}$, $W_{xf}$, $W_{xo}$ and $W_{xg}$ are the weight matrices of each of above four layers connecting to the input vector.
$W_{hi}$, $W_{hf}$, $W_{ho}$, and $W_{hg}$ are the weight matrices of each of layers connecting to the previous short-term state.
$b_i$, $b_f$, $b_o$, and $b_g$ are the bias terms for each of layer.

In a deep neural network, the training may suffer from overfitting due to a large number of its own hyperparameters. We can use the method called $regularization$ to prevent it from overfitting. $Dropout$ is one of the most popular $regularization$ techniques, applying in some layers to  reduce the overfitting risk \citep{Aurelien:2017}. In this way, some neurons has a probability of being ignored at every step controlled by $dropout \ rate$. Besides, it is also of benefit to estimate the confidence of the training in BNN.

BNN is a supplementary of RNN for calculating the uncertainty of the prediction. BNN is defined in terms of a prior distribution with parameters over the wights $p(\omega)$, which manifests a prior belief about parameters generating the observations. With a given dataset $\{\textbf{X},\textbf{Y}\}$, we can achieve the posterior distribution of the parameters space $p(\omega |\textbf{X},\textbf{Y})$. Thus the output of a new input point $x$ can be anticipated by the integration
\begin{equation}
p(y^*|x^*,\textbf{X},\textbf{Y})=\int p(y^*|x^*,\omega)p(\omega|\textbf{X},\textbf{Y})\textrm{d}\omega.
\end{equation}

A full BNN is extremely complex. Several works had shown that a network with a dropout is approximately equivalent to the Bayesian model \citep{Gal:2016a,Gal:2016b,Gal:2016c,Louizos:2016}. Introducing the Bayesian machinery into the deep learning framework, \citet{Gal:2016a,Gal:2016b,Gal:2016c} developed a new framework casting dropout training in deep neural network as approximate Bayesian inference in deep Gaussian processes and successfully applied in RNN. Their results offer a Bayesian interpretation of the dropout technique, and verify that a network with a dropout is mathematically equivalent to the Bayesian model. When the RNN is well-trained and executed $n$ times, the network is equivalent to BNN. Therefore, we employ the $dropout$ in the training and call the trained network $n$ times to estimate the uncertainty of outputs, where the $dropout$ is an approximation of the Gaussian processes and cooperates with the activation function to determine the confidence regions of prediction.

\section{Reconstructing the distance-redshift relation from Pantheon}\label{sec:Reconstruction}

In order to reconstruct the Hubble diagram to high redshift range with our network, we use the latest Pantheon compilation \citep{Scolnic:2018} of SNe Ia as the training data. The Pantheon compilation consists of 1048 well-calibrated SNe Ia in the redshift range $0.01<z<2.3$. The distance modulus of SNe Ia is given by
\begin{equation}\label{eq:mu}
\mu_B(z;\alpha,\beta,M_b)=m-M_b+\alpha x(z)-\beta c(z),
\end{equation}
where $m$ is the apparent magnitude, $M_b$ is the absolute magnitude, $x$ and $c$ are the stretch factor and color parameter respectively, $\alpha$ and $\beta$ are the nuisance parameter. In the Pantheon dataset, the presented apparent magnitude has already been corrected for stretch and color, thus the stretch and color corrections are vanishing in eq.(\ref{eq:mu}). The absolute magnitude is fixed to $M_b=-19.36$. We obtain the distance moduli $\mu(z)$ from eq.(\ref{eq:mu}) and sort the data points according to the redshift from low to high.

We construct the RNN+BNN network and train it with the package TensorFlow\footnote{https://www.tensorflow.org}. For clarity, we present the corresponding hyperparameters in Figure \ref{fig_ourRNN} and list the steps to reconstruct data with our network as follow: (a) Data processing. The scale of data has an effect on training. Hence, we normalize the distance moduli of the sorted Pantheon data and re-arrange $\mu-z$ as sequences with the step number $t=4$. (b) Building RNN. We build RNN with three layers, i.e. an input layer, a hidden layer and an output layer as described in Figure \ref{fig_ourRNN}. The first two layers are constructed with the LSTM cells of 100 neurons. The redshifts $z^{<t>}$ and the corresponding distance moduli $\mu^{<t>}$ are the input and output vectors, respectively. We employ the Adam optimizer to minimize the cost function MSE and train the network 1000 times. (c) Building BNN. We set the dropout rate to 0 in the input layer to avoid the lost of information, and to 0.2 in the second layer as well as the output layer \citep{Bonjean:2020,Mangena:2020}. We execute the trained network 1000 times to obtain the distribution of distance moduli.

\begin{figure}[htbp]
  \centering
  \includegraphics[width=0.8\textwidth]{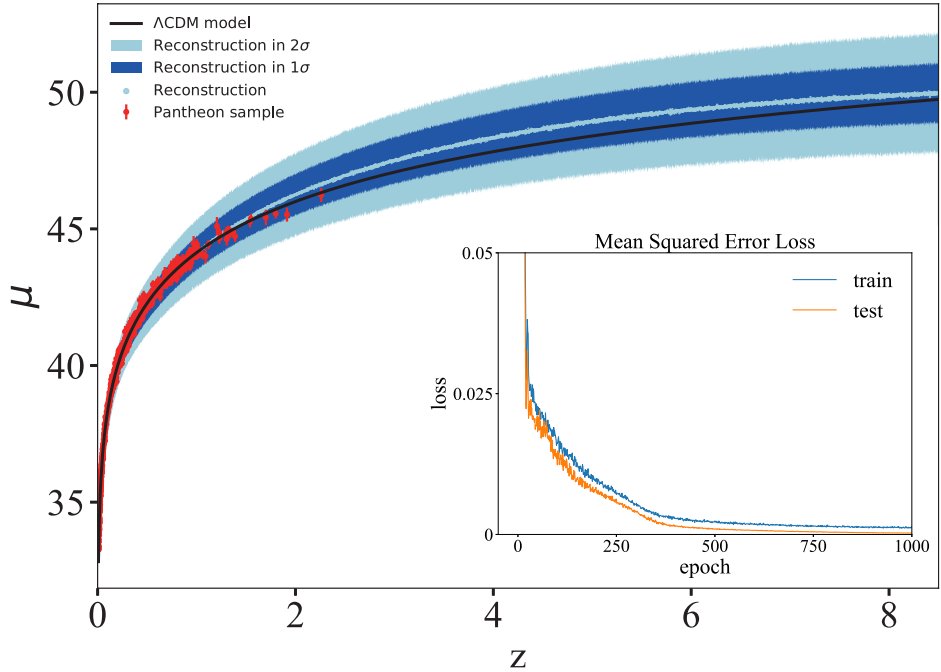}
  \caption{\small{The reconstruction of distance moduli from Pantheon data set. The red dots with $1\sigma$ error bars are the Pantheon data points. The light-blue dots are the central values of reconstruction. The shaded regions are the 1$\sigma$ and 2$\sigma$ uncertainties. The black curve is the best-fitting curve of flat $\Lambda$CDM model. Inside plot: the Mean Squared Error (MSE) loss over the training epochs for both the train (blue) and test (orange) sets.}}\label{fig_simulation}
\end{figure}

The highest redshift of GRB sample \citep{Wang:2011} is 8.2. Therefore, we reconstruct the distance moduli by training RNN+BNN up to $z=8.5$. The result is shown in Figure \ref{fig_simulation}. The red dots with error bars and the light-blue dots are the observational data points of Pantheon and the central values of reconstruction, respectively. The shaded regions are the 1$\sigma$ and 2$\sigma$ uncertainties of the reconstruction. The inside plot at the lower-right corner of Figure \ref{fig_simulation} is the MSE loss over the training epochs for both the train (blue) and test (orange) sets, which approach their minimum value when epoch near 400. For comparison, we also plot the best-fitting curve of the $\Lambda$CDM model (black line). The result shows that, our reconstruction is excellently consistent with the $\Lambda$CDM model within 1$\sigma$ confidence level. We want to emphasize again that our reconstruction of distance moduli is neither dependent on the cosmological model, nor has any assumption about the concrete form of the reconstructing function. It is the architecture itself makes the neural network become a universal approximator, not the specific activation function \citep{Hornik:1991}. We note that the Gaussian processes can also reconstruct the distance moduli without involving any model assumption \citep{Lin:2018mdj}. However, the reconstructed uncertainty of Gaussian processes is very large in the region where the data points are sparse. The advantage of RNN+BNN compared with the Gaussian processes is that, the former can reconstruct the curve precisely even far beyond the data points.

\begin{figure}[htbp]
  \centering
  \includegraphics[width=0.48\textwidth]{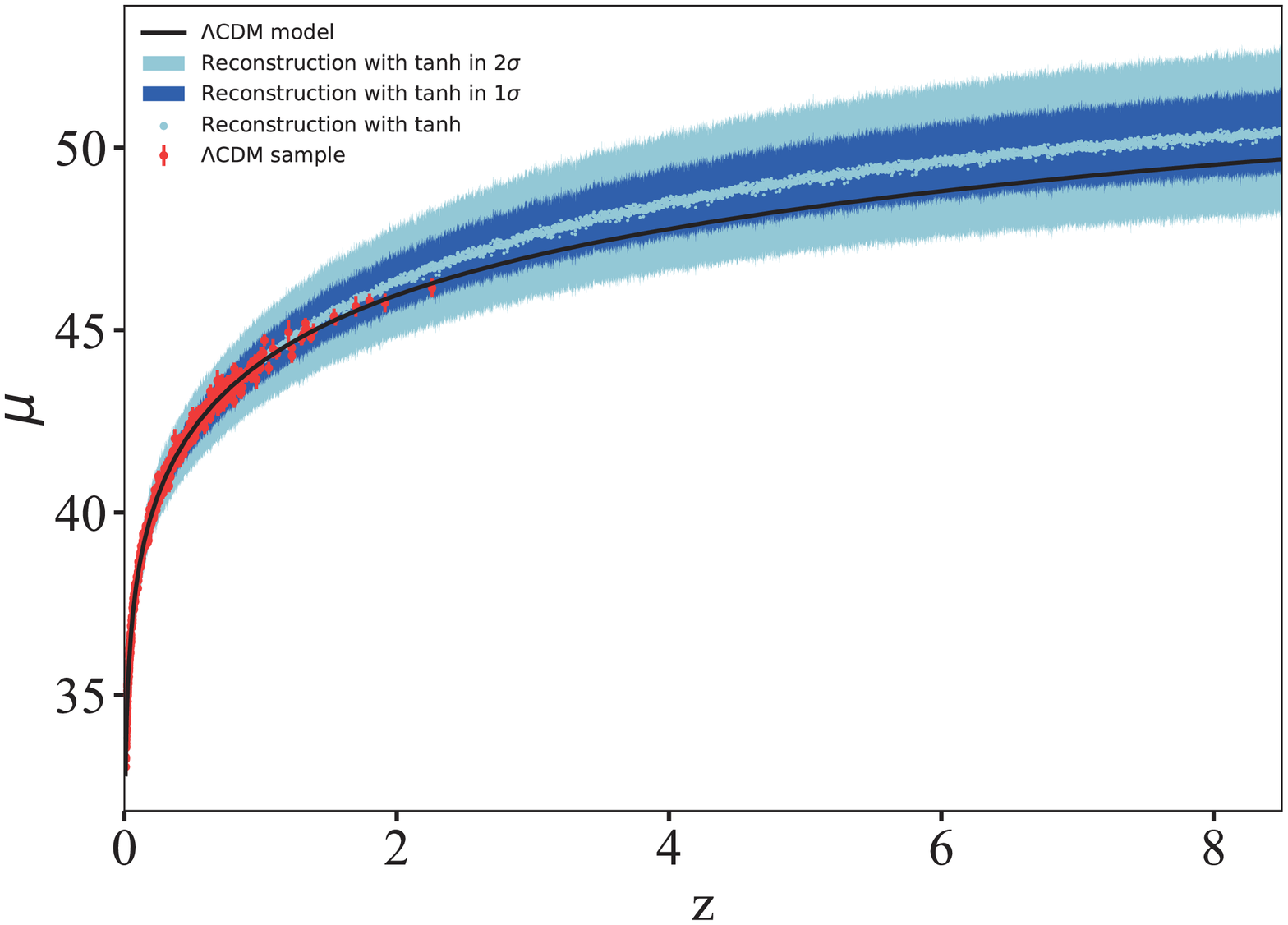}
  \includegraphics[width=0.48\textwidth]{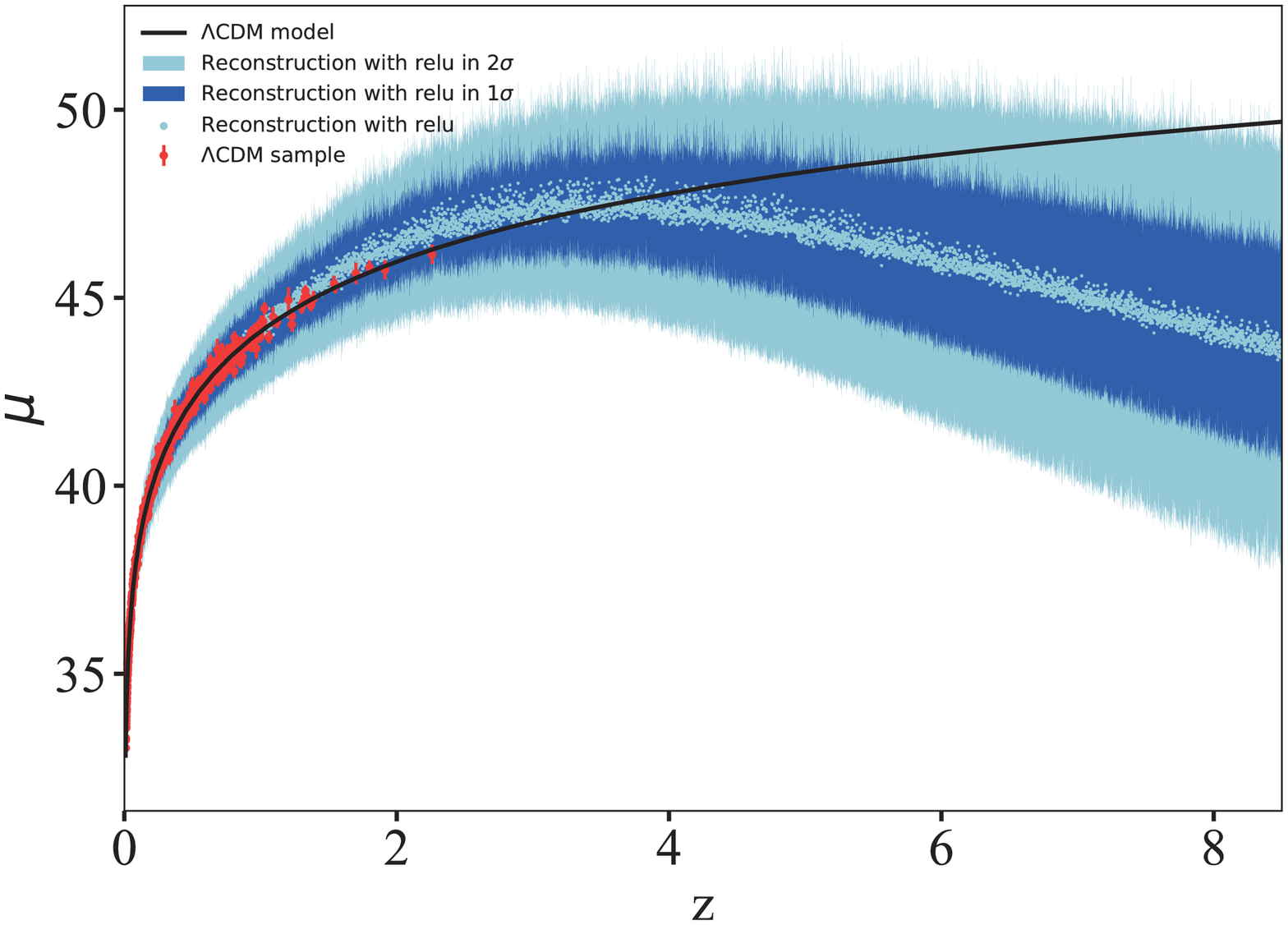}
  \includegraphics[width=0.48\textwidth]{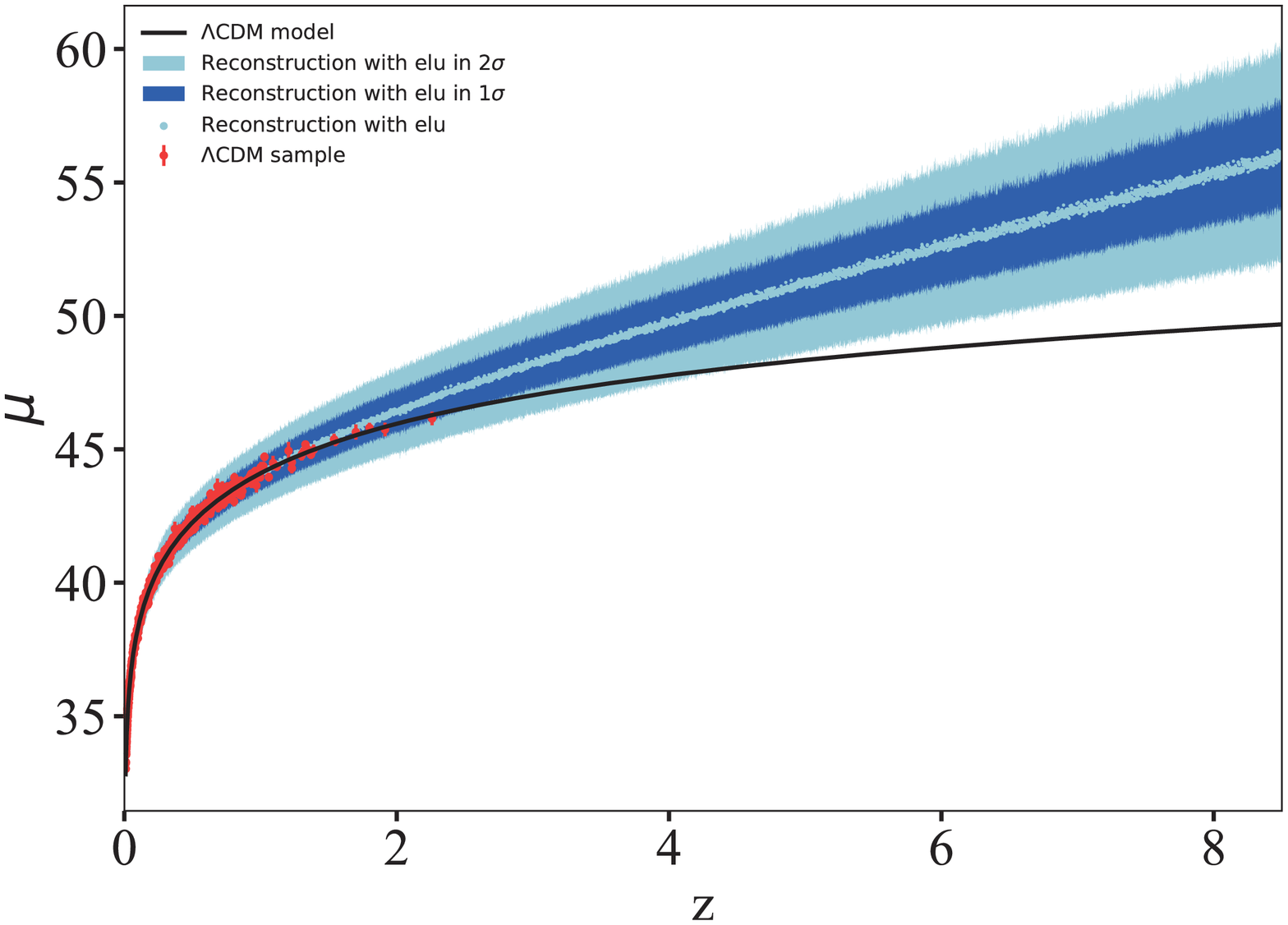}
  \includegraphics[width=0.48\textwidth]{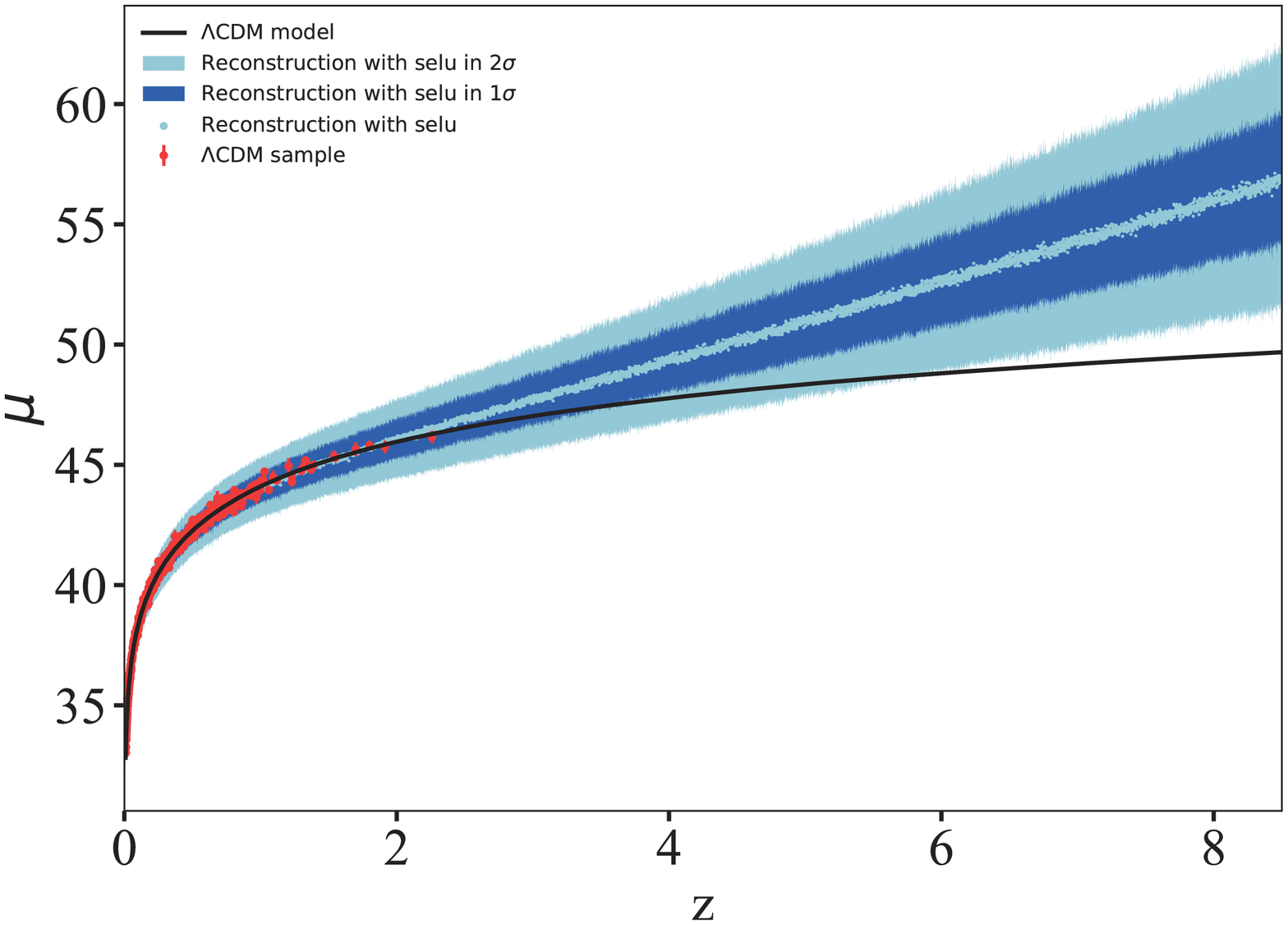}
  \caption{\small{The reconstructions of distance-redshift relation from Pantheon-like sample (generated from fiducial $\Lambda$CDM model) with four different activation functions. Top-left: tanh; top-right: relu; bottom-left: elu; bottom-right: selu. This figure shows that the reconstruction depends on the activation functions.}}\label{fig_Af_compare}
\end{figure}

To investigate the influence of different choices of activation function on the reconstruction, we use Monte Carlo simulations. We first simulate a set of Pantheon-like sample, whose redshifts and distance uncertainties are the same to that of Pantheon sample, and the distance moduli are sampled from Gaussian distribution $\mu\sim G(\bar\mu,\sigma_\mu)$, where $\bar\mu$ is calculated using the fiducial $\Lambda$CDM model with $\Omega_M=0.3$ and $H_0=70~{\rm km/s/Mpc}$, and $\sigma_\mu$ is the uncertainty of the observed distance modulus of Pantheon sample. Then we replace the Pantheon data with the mock data and retrain the network, and then reconstruct the distance-redshift relation using the trained network. We also consider other three activation functions as a comparison with tanh function:
\begin{equation}
A_{f_{\rm Relu}}=
\begin{cases}
0,& x\leq0\\
x,& x>0
\end{cases}
\end{equation}
\begin{equation}
A_{f_{\rm Elu}}=
\begin{cases}
\alpha\left(e^x-1\right),& x\leq0\\
x,& x>0
\end{cases}
\end{equation}
\begin{equation}
A_{f_{\rm Selu}}=
\begin{cases}
\alpha\lambda\left(e^x-1\right),& x\leq0\\
\lambda x,& x>0
\end{cases}
\end{equation}
where in Elu function $\alpha=1$ \citep{Clevert:2015}, and in Selu function $\alpha$ $\approx$ 1.673 and $\lambda$ $\approx$ 1.051 \citep{Klambauer et al.:2017}. The reconstructed results with four different activation functions are presented in Figure \ref{fig_Af_compare}. It is shown that only with the tanh function our network can correctly reconstruct the distance-redshift relation up to $z\approx 8.5$ within $1\sigma$ uncertainty. For the rest three activation functions, the network couldn't correctly reconstruct the curve at high redshift. Therefore, we choose the tanh function rather than the rest three in our work. Note that the similarity between the elu and selu functions makes the reconstructed curves be similar to each other.

\begin{figure}[htbp]
  \centering
  \includegraphics[width=0.8\textwidth]{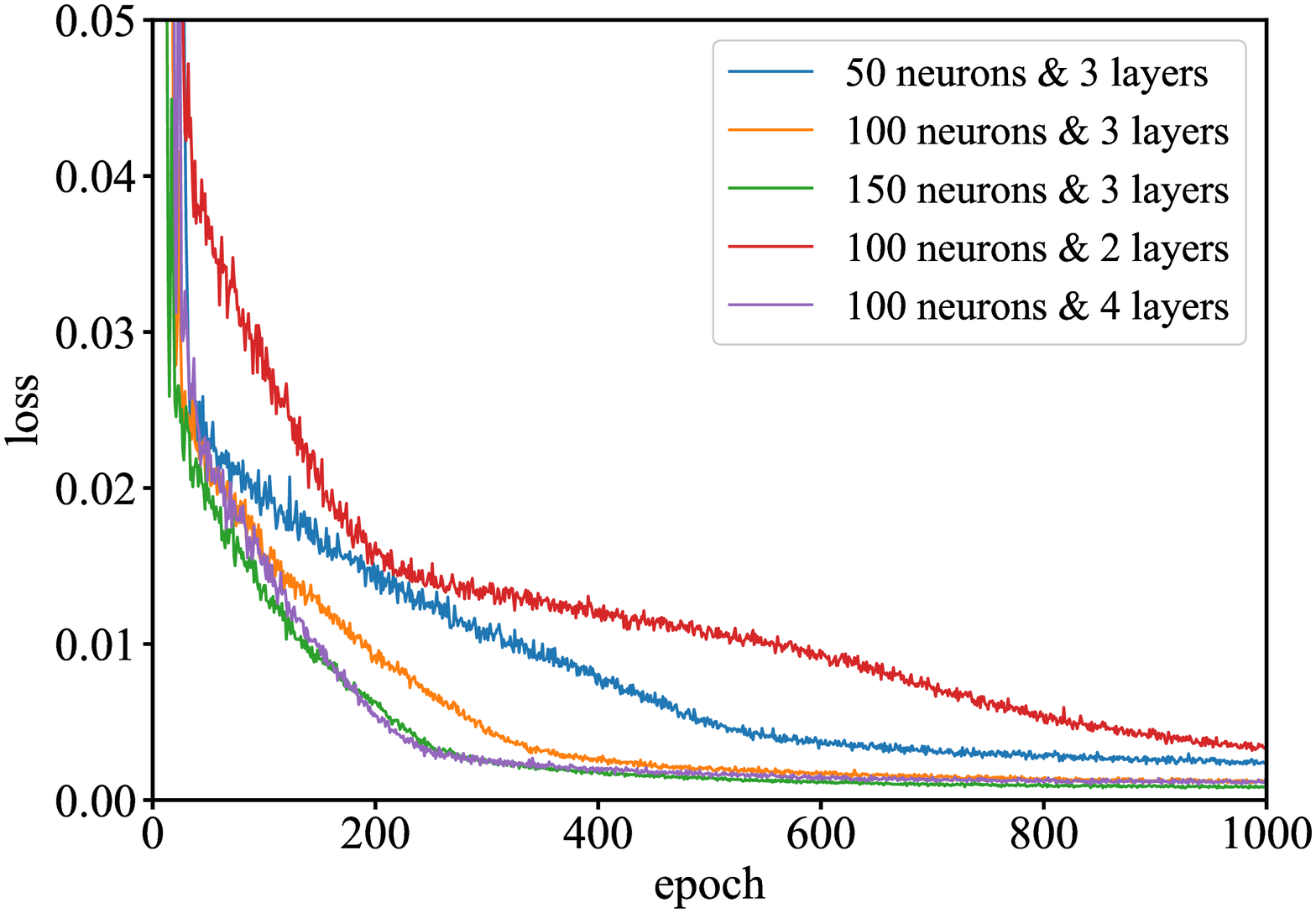}
  \caption{\small{The training loss in the cases of different numbers of neurons and layers with the activation function fixed to tanh.}}\label{fig_par_compare}
\end{figure}

We fix the activation function to tanh to investigate the influence of other hyperparameters on our network. Firstly, we fix the number of layers to 3 and choose three different number of neurons \{50, 100 150\} for comparison. We also fix the number of neurons to 100 and choose three different number of layers \{2,3,4\} for comparison. The training losses of different cases are presented in Figure \ref{fig_par_compare}. For the same number of layers, the training loss of 50 neurons is distinctly higher than that of 100 and 150 neurons. The increase of neurons from 100 to 150 does not significantly reduce the training loss. For the same number of neurons, the training loss of 2 layers is much higher than that of 3 and 4 layers, but the training losses of 3 layers and 4 layers are not significantly distinguishable. Therefore, our choice of 100 neurons and 3 layers is appropriate to construct a well-trained network.

\section{Testing the redshift dependence of luminosity correlations}\label{sec:Test}

After reconstructing the distance-redshift relation from SNe, we can use it to calibrate the luminosity correlations of GRBs. The luminosity correlations of GRB can be expressed with the exponential form $R=AQ^b$, and further be re-expressed with the linear form by taking logarithms,
\begin{equation}
y=a+bx  \ \ \ (y\equiv \log R, x\equiv \log Q, a\equiv \log A),
\end{equation}
in which `log' denotes the logarithm of base 10. We definitely write the six luminosity correlations as follows
\begin{equation}
\log\frac{L}{\textrm{erg} \ \textrm{s}^{-1}}=a_1+b_1\log\frac{\tau_{\textrm{lag},i}}{0.1s},
\end{equation}
\begin{equation}
\log\frac{L}{\textrm{erg} \ \textrm{s}^{-1}}=a_2+b_2\log\frac{V_i}{0.02},
\end{equation}
\begin{equation}
\log\frac{L}{\textrm{erg} \ \textrm{s}^{-1}}=a_3+b_3\log\frac{E_{p,i}}{300\textrm{keV}},
\end{equation}
\begin{equation}\label{eq:Ep-Egamma}
\log\frac{E_{\gamma}}{\textrm{erg} }=a_4+b_4\log\frac{E_{p,i}}{300\textrm{keV}},
\end{equation}
\begin{equation}
\log\frac{L}{\textrm{erg} \ \textrm{s}^{-1}}=a_5+b_5\log\frac{\tau_{\textrm{RT},i}}{0.1\textrm{s}},
\end{equation}
\begin{equation}
\log\frac{E_{\textrm{iso}}}{\textrm{erg}}=a_6+b_6\log\frac{E_{p,i}}{300\textrm{keV}}.
\end{equation}
Here, quantities with a subscript `\textit{i}' indicate that they are measured in the comoving frame, which are related to the quantities in the observer frame by $\tau_{\textrm{lag},i}=\tau_{\textrm{lag}}(1+z)^{-1}$, $\tau_{\textrm{RT},i}=\tau_{\textrm{RT}}(1+z)^{-1}$, $V_i=V(1+z)$ and $E_{p,i}=E_p(1+z)$, where ${\tau_{\textrm{lag}},\tau_{\textrm{RT}},V,E_p}$ can be directly derived from the observations of spectrum or light curve of GRBs.

Assuming that GRBs radiate isotropically, the isotropic equivalent luminosity can be derived from the bolometric peak flux $P_{\textrm{bolo}}$ by \citep{Schaefer:2007}
\begin{equation}
L=4\pi d^2_L P_{\textrm{bolo}},
\end{equation}
where $d_L$ is the luminosity distance of GRB, which can be obtained from the reconstructed distance moduli of Pantheon presented in section \ref{sec:Reconstruction} with the relation
\begin{equation}\label{eq:mu-DL}
\mu=5\log\frac{d_L}{\textrm{Mpc}}+25.
\end{equation}
Hence, the uncertainty of $L$ propagates from the uncertainties of $P_{\rm bolo}$ and $d_L$. The isotropic equivalent energy $E_{\rm iso}$ can be obtained from the bolometric fluence $S_{\rm bolo}$ by
\begin{equation}
E_{\rm iso}=4\pi d^2_L S_{\textrm{bolo}} (1+z)^{-1},
\end{equation}
the uncertainty of $E_{\rm iso}$ propagates from the uncertainties of $S_{\rm bolo}$ and $d_L$. If, on the other hand, GRBs radiate in two symmetric beams, then we can define the collimation-corrected energy $E_{\gamma}$ as
\begin{equation}\label{eq:Egamma-DL}
E_{\gamma}\equiv E_{\textrm{iso}}F_{\textrm{beam}},
\end{equation}
where $F_{\rm beam}\equiv 1-\cos \theta_{\rm jet}$ is the beaming factor, $\theta_{\rm jet}$ is the jet opening angle. The uncertainty of $E_{\gamma}$ propagates from the uncertainties of $E_{\rm iso}$ and $F_{\rm beam}$.

Our GRBs sample is taken from \cite{Wang:2011}, which consists of 116 long GRBs in redshift range $z\in[0.17,8.2]$. Following \cite{Lin:2016}, we divide GRBs into two subsamples, i.e. the low-$z$ sample ($z\leq 1.4$) which consists of 50 GRBs, and the high-$z$ sample ($z> 1.4$) which consists of 66 GRBs. We investigate the redshift dependence of luminosity correlations for this two subsamples, as well as for the full GRBs sample. To fit the six luminosity correlations, we apply the D'Agostini's liklihood \citep{DAgostini:2005}
\begin{equation}
\mathcal{L}\left(\sigma_{\rm int},a,b\right)\propto\prod_i\frac{1}{\sqrt{\sigma^2_{\rm int}+\sigma^2_{yi}+b^2\sigma^2_{xi}}}\times \exp\left[-\frac{(y_i-a-bx_i)^2}{2(\sigma^2_{\rm int}+\sigma^2_{yi}+b^2\sigma^2_{xi})}\right].
\end{equation}
By maximizing this joint likelihood function, we can derive the best-fitting parameters $(a,b,\sigma_{\rm int})$, where the intrinsic scatter $\sigma_{\rm int}$ denotes any other unknown errors except for the measurement errors.

\begin{table}[htbp]
\centering
\caption{\small{The best-fitting parameters of GRB luminosity correlations. $N$ is the number of GRBs in each subsample.}}\label{tab:parameters}
\arrayrulewidth=1.0pt
\renewcommand{\arraystretch}{1.3}
{\begin{tabular}{cccccc} 
\hline\hline 
Correlation &Sample & N  & $a$     & $b$   &$\sigma_{\rm int}$\\\hline

$\tau_{\rm lag}-L$ &low-z  &37 &52.076$\pm$0.106& 	-0.789$\pm$0.159& 0.495$\pm$0.0888\\
               &high-z &32 &52.610$\pm$0.069& 	-0.633$\pm$0.115& 	0.144$\pm$0.099\\
               &All-z &69 &52.313$\pm$0.073& 	-0.757$\pm$0.114& 	0.462$\pm$0.059\\
\hline
$V-L$ &low-z  &47 &51.524$\pm$0.202& 	0.440$\pm$0.301& 	0.863$\pm$0.084\\
      &high-z &57 &52.355$\pm$0.131& 	0.264$\pm$0.125& 	0.480$\pm$0.073\\
      &All-z &104 &51.784$\pm$0.140& 	0.558$\pm$0.160& 	0.747$\pm$0.070\\
\hline
$E_p-L$ 	&low-z  &50 &51.855$\pm$0.092& 	1.481$\pm$0.185& 	0.576$\pm$0.075\\
            &high-z &66 &52.495$\pm$0.053& 	1.151$\pm$0.139& 	0.192$\pm$0.088\\
            &All-z &116 &52.167$\pm$0.059& 	1.454$\pm$0.136& 	0.530$\pm$0.049\\
\hline
$E_p-E_{\gamma}$ &low-z  &12  &50.602$\pm$0.088& 	1.538$\pm$0.192& 	$<$0.176\\
                 &high-z &12  &50.768$\pm$0.126& 	1.162$\pm$0.377& 	$<$0.238\\
                 &All-z &24   &50.650$\pm$0.067& 	1.486$\pm$0.157& 	$<$0.137\\
\hline
$\tau_{\rm RT}-L$ &low-z  &39  &52.671$\pm$0.132& 	-1.354$\pm$0.194& 0.458$\pm$0.074\\
              &high-z &40  &52.880$\pm$0.080& 	-0.816$\pm$0.161& 	0.249$\pm$0.097\\
              &All-z &79   &52.766$\pm$0.078& 	-1.250$\pm$0.132& 	0.424$\pm$0.054\\
\hline
$E_p-E_{\rm iso}$ &low-z  &40 &52.536$\pm$0.099& 	1.600$\pm$0.201& 0.557$\pm$0.085\\
              &high-z &61 &53.013$\pm$0.055& 	1.293$\pm$0.135& 	0.192$\pm$0.088\\
              &All-z &101 &52.778$\pm$0.058& 	1.546$\pm$0.129& 	0.461$\pm$0.051\\
\hline
\end{tabular}}
\end{table}

\begin{figure}[htbp]
\centering
\includegraphics[width=0.48\textwidth]{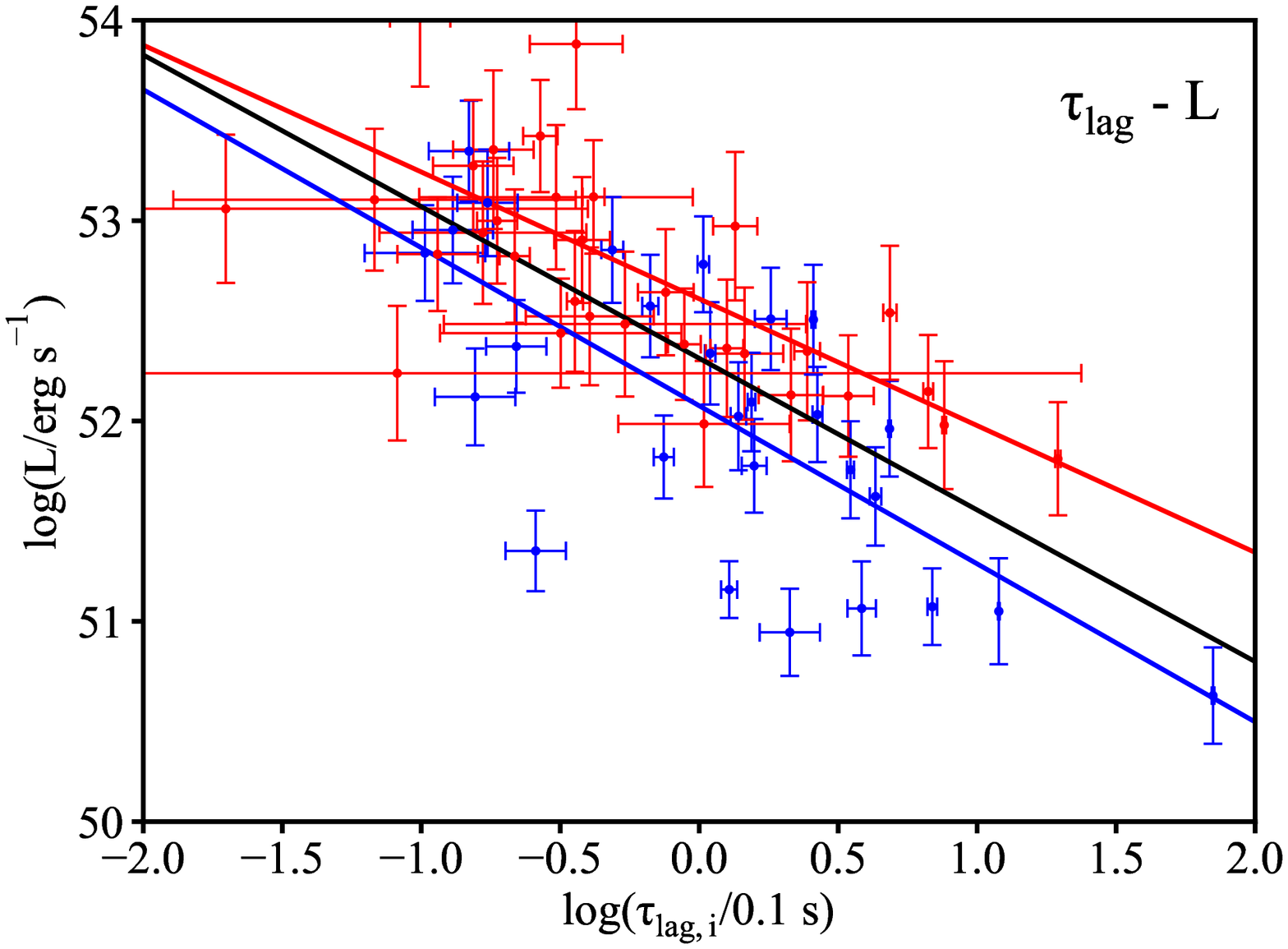}
\includegraphics[width=0.48\textwidth]{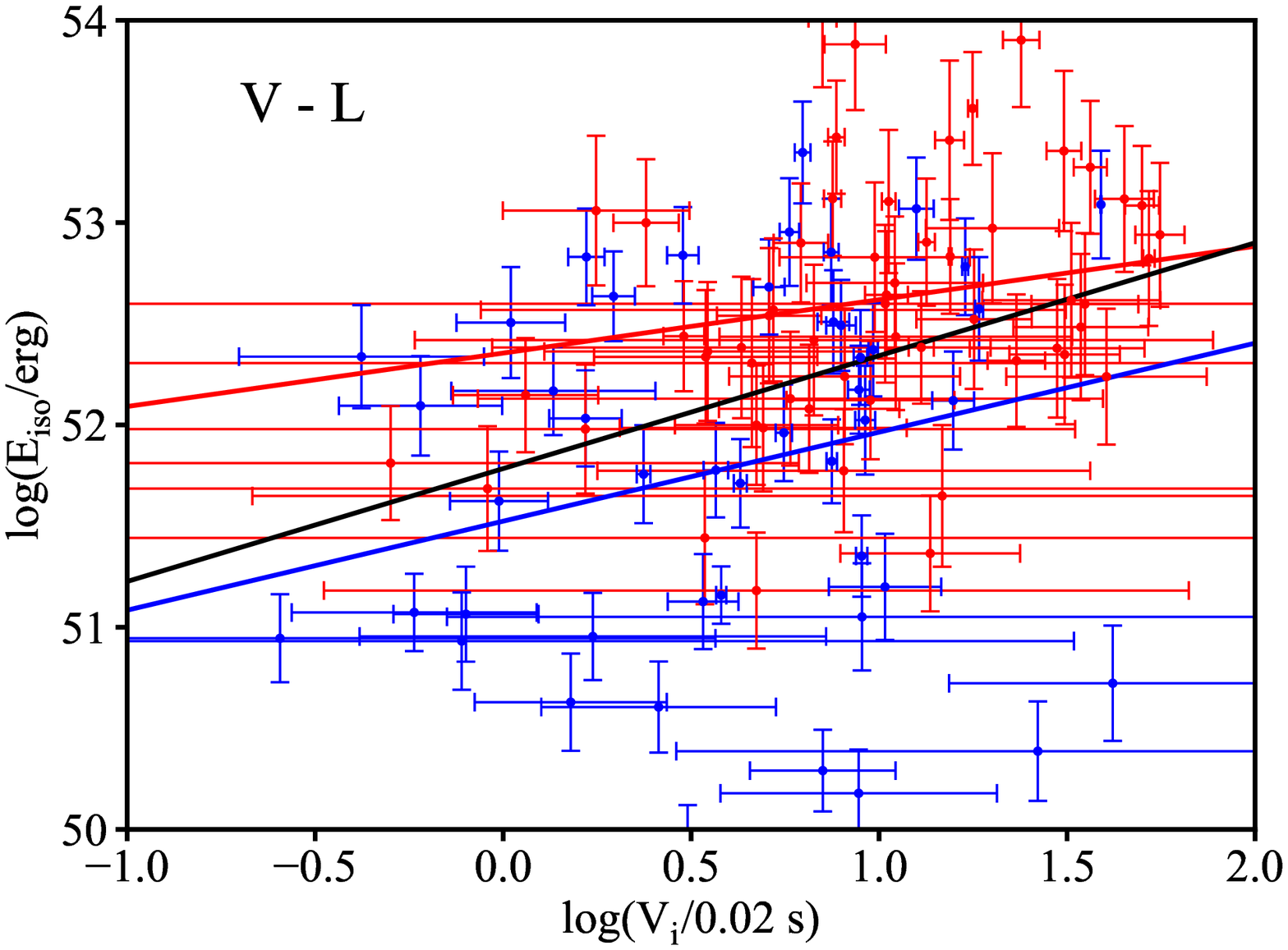}
\includegraphics[width=0.48\textwidth]{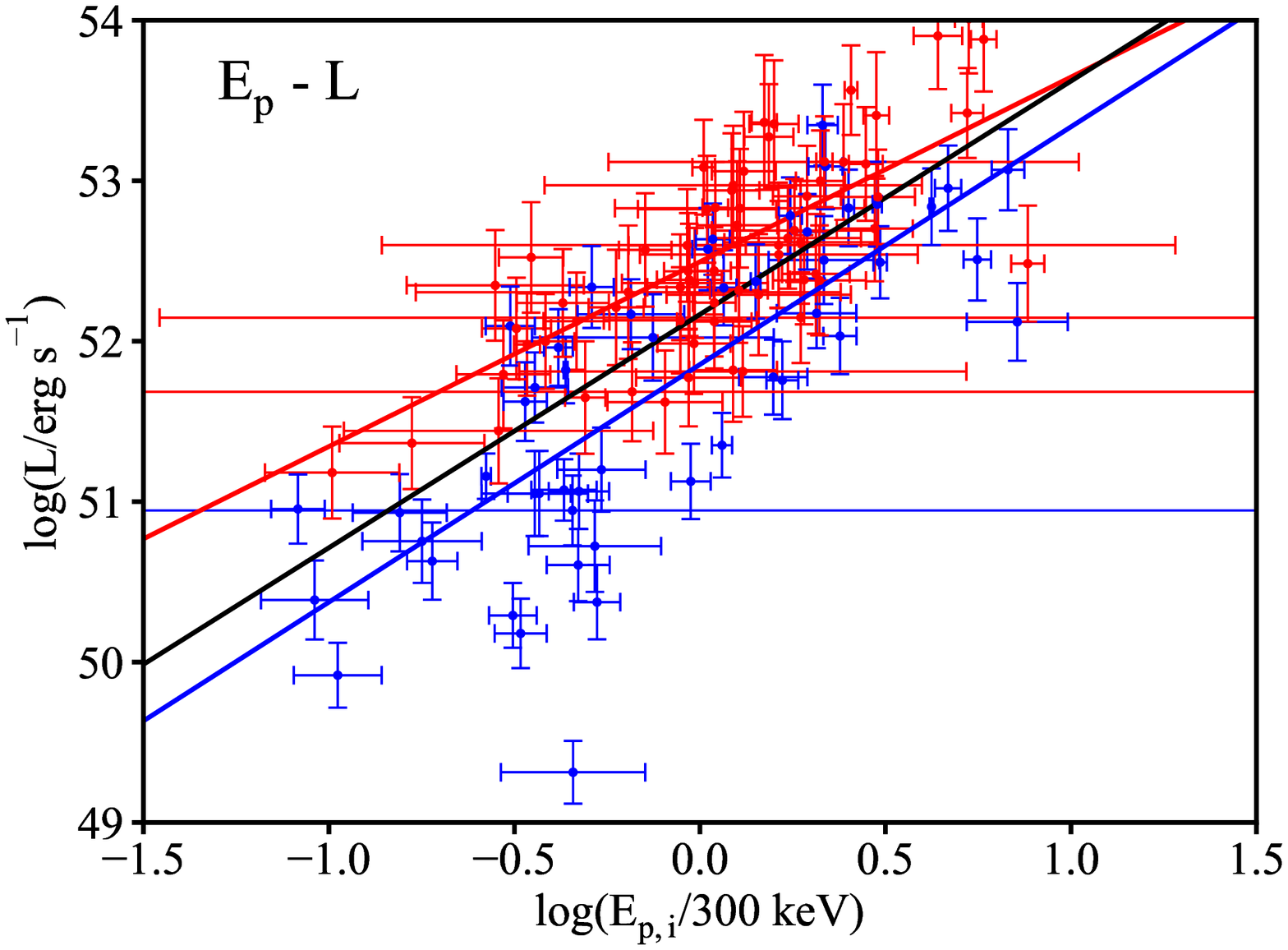}
\includegraphics[width=0.48\textwidth]{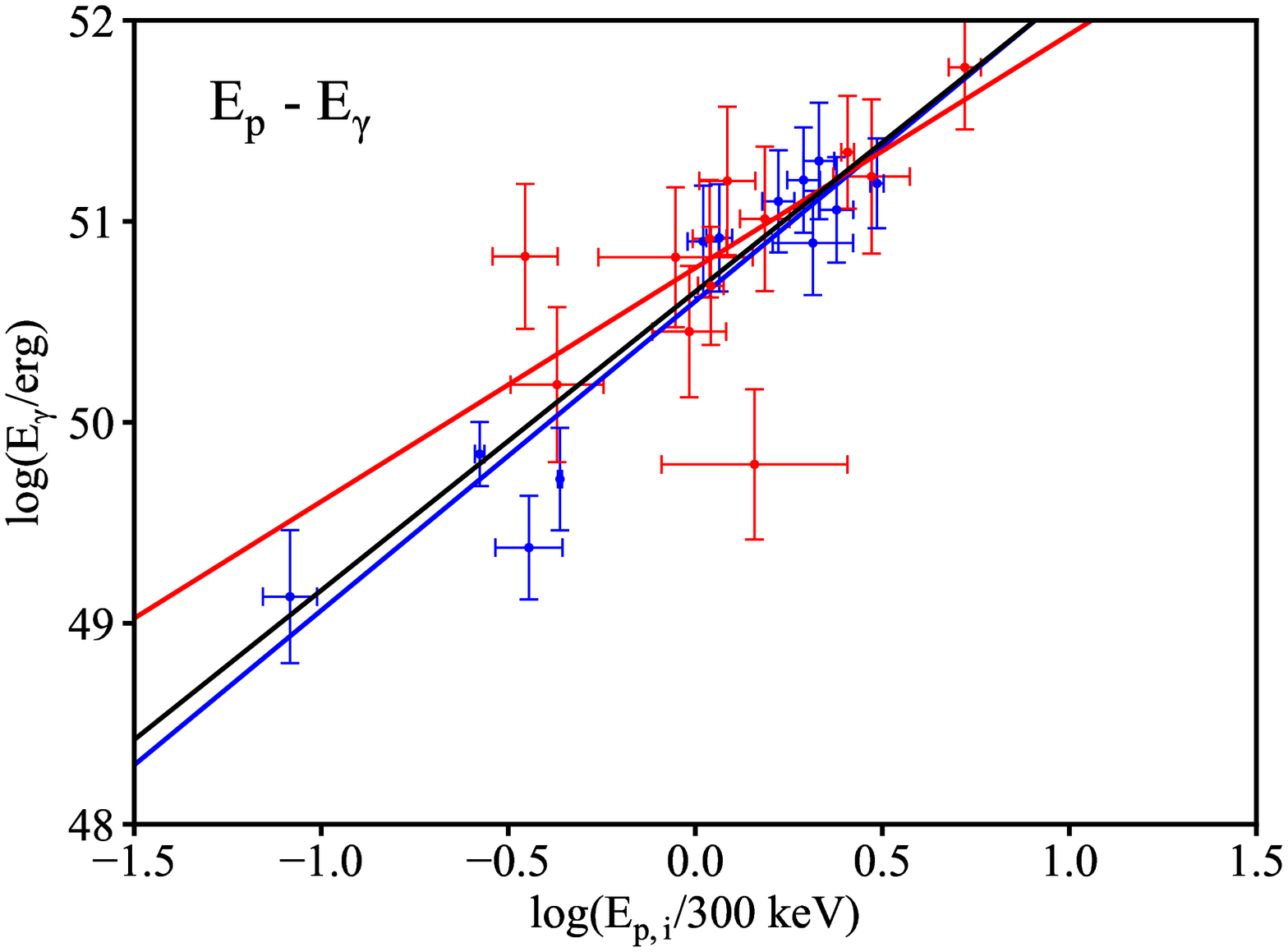}
\includegraphics[width=0.48\textwidth]{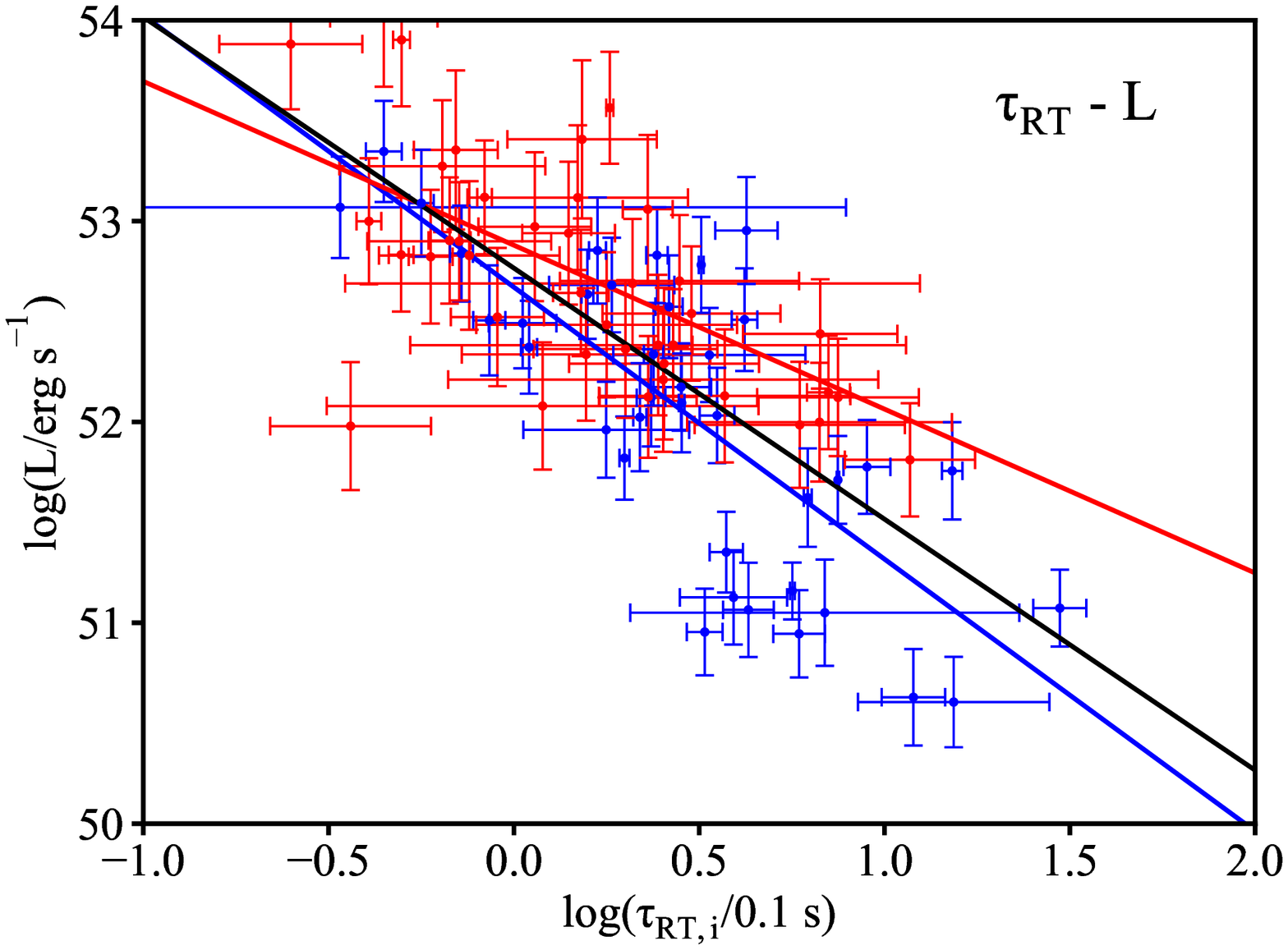}
\includegraphics[width=0.48\textwidth]{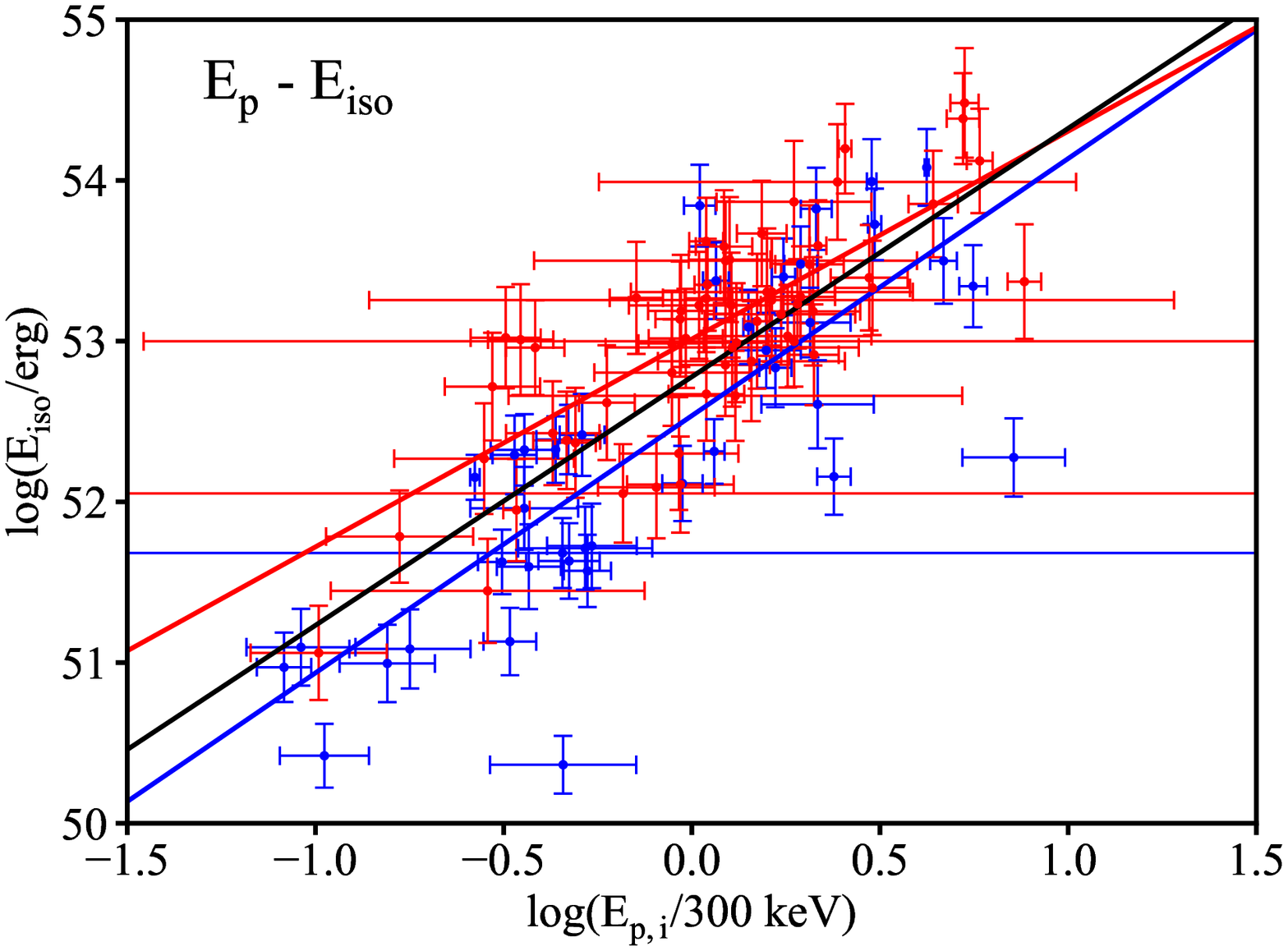}
\caption{\small{The luminosity correlations for low-$z$ (blue dots) and high-$z$ (red dots) GRBs. Error bars denote the 1$\sigma$ uncertainties. The lines are the best-fitting results, blue line for low-$z$ GRBs, red line for high-$z$ GRBs and black line for all-$z$ GRBs.}}\label{fig_xy}
\end{figure}

\begin{figure}[htbp]
\centering
\includegraphics[width=0.44\textwidth]{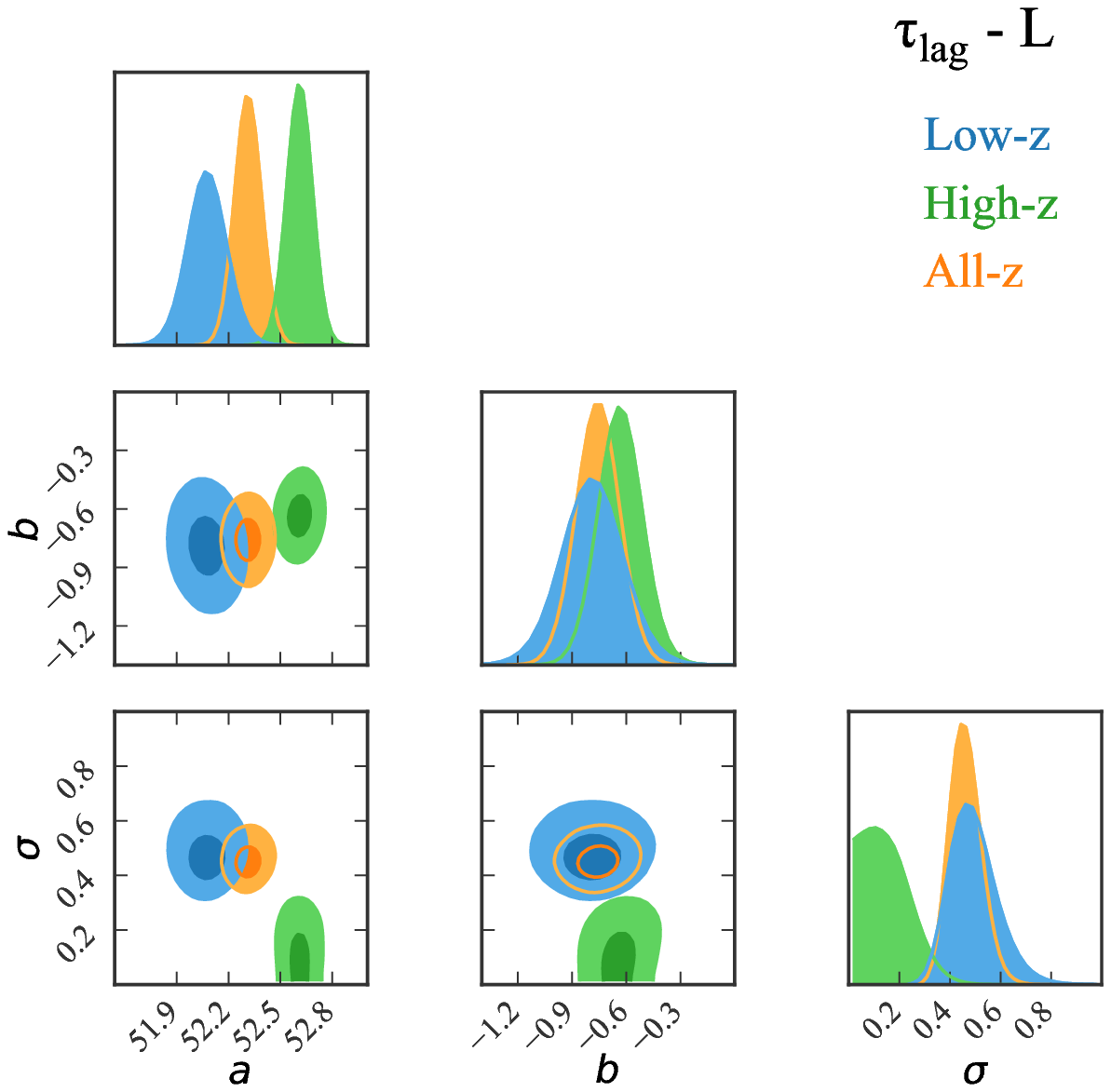}
\includegraphics[width=0.44\textwidth]{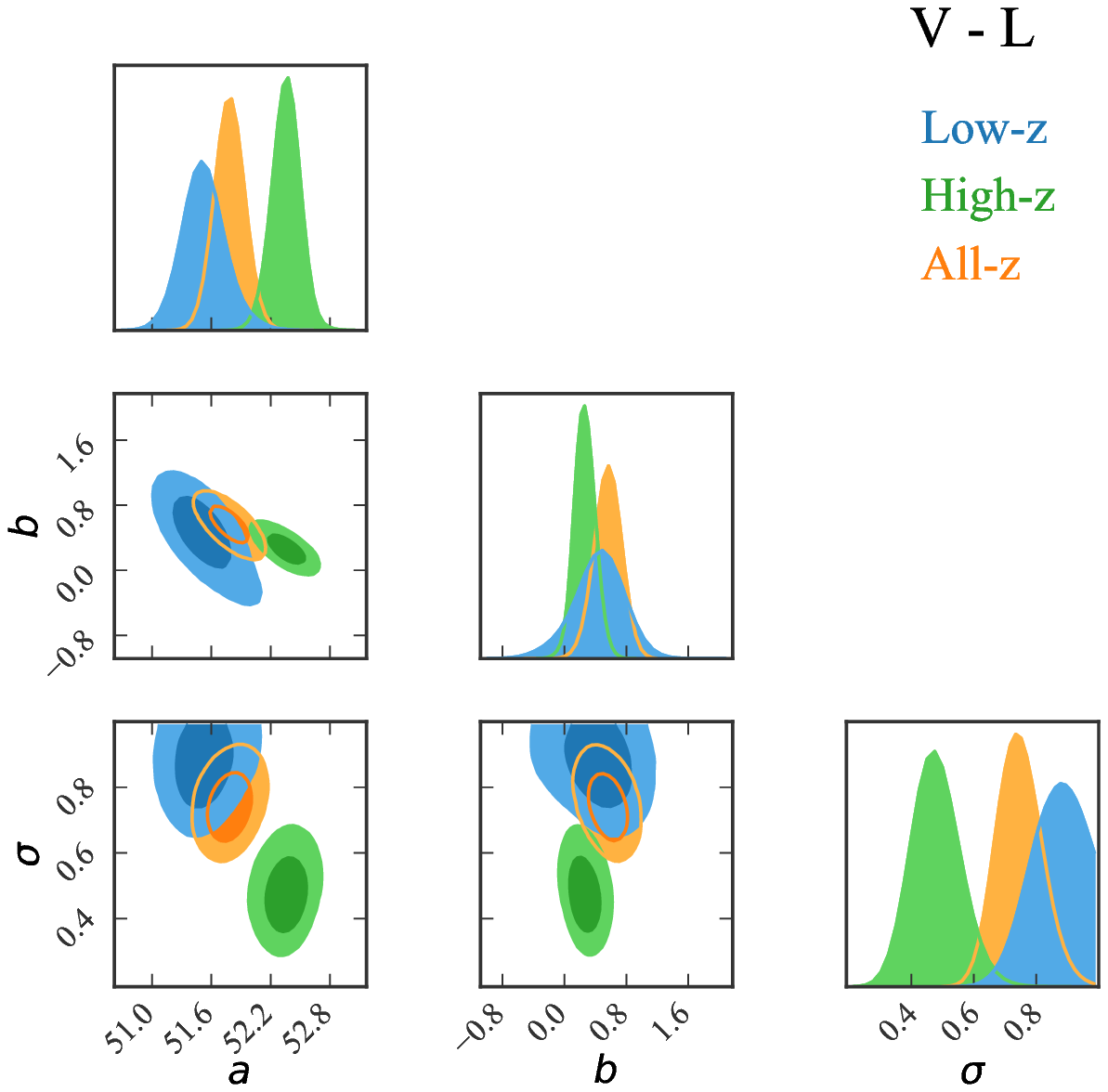}
\includegraphics[width=0.44\textwidth]{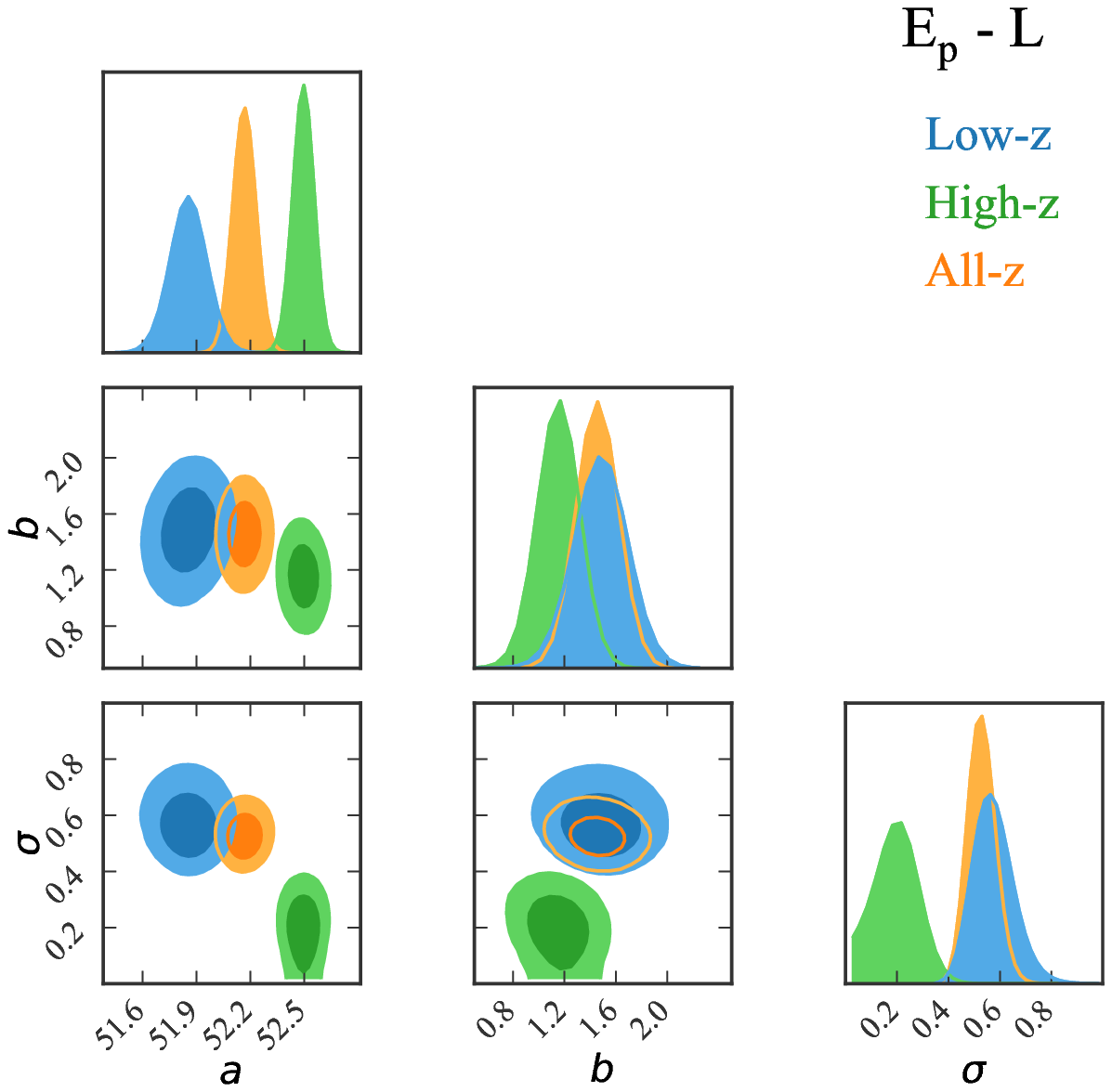}
\includegraphics[width=0.44\textwidth]{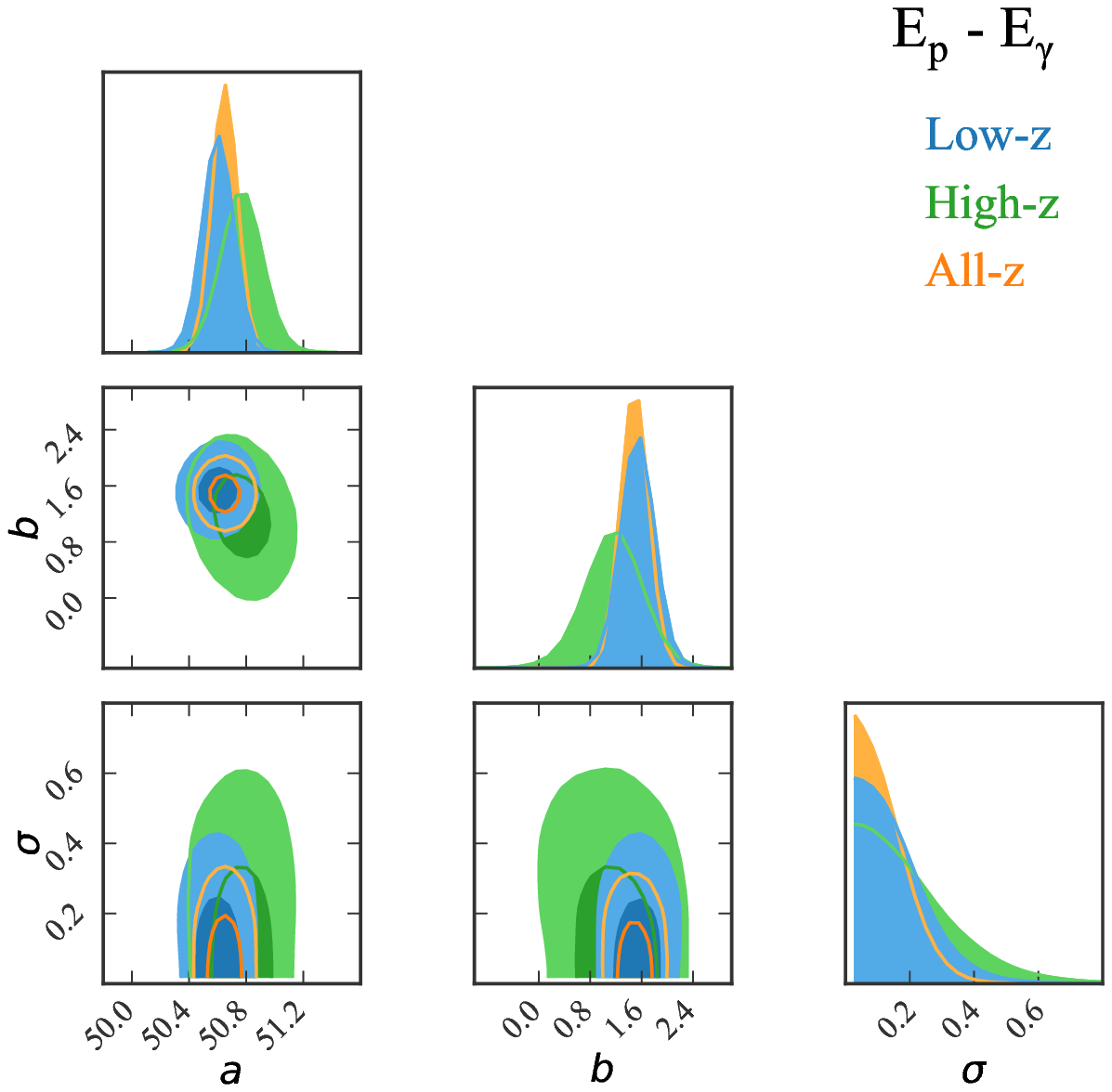}
\includegraphics[width=0.44\textwidth]{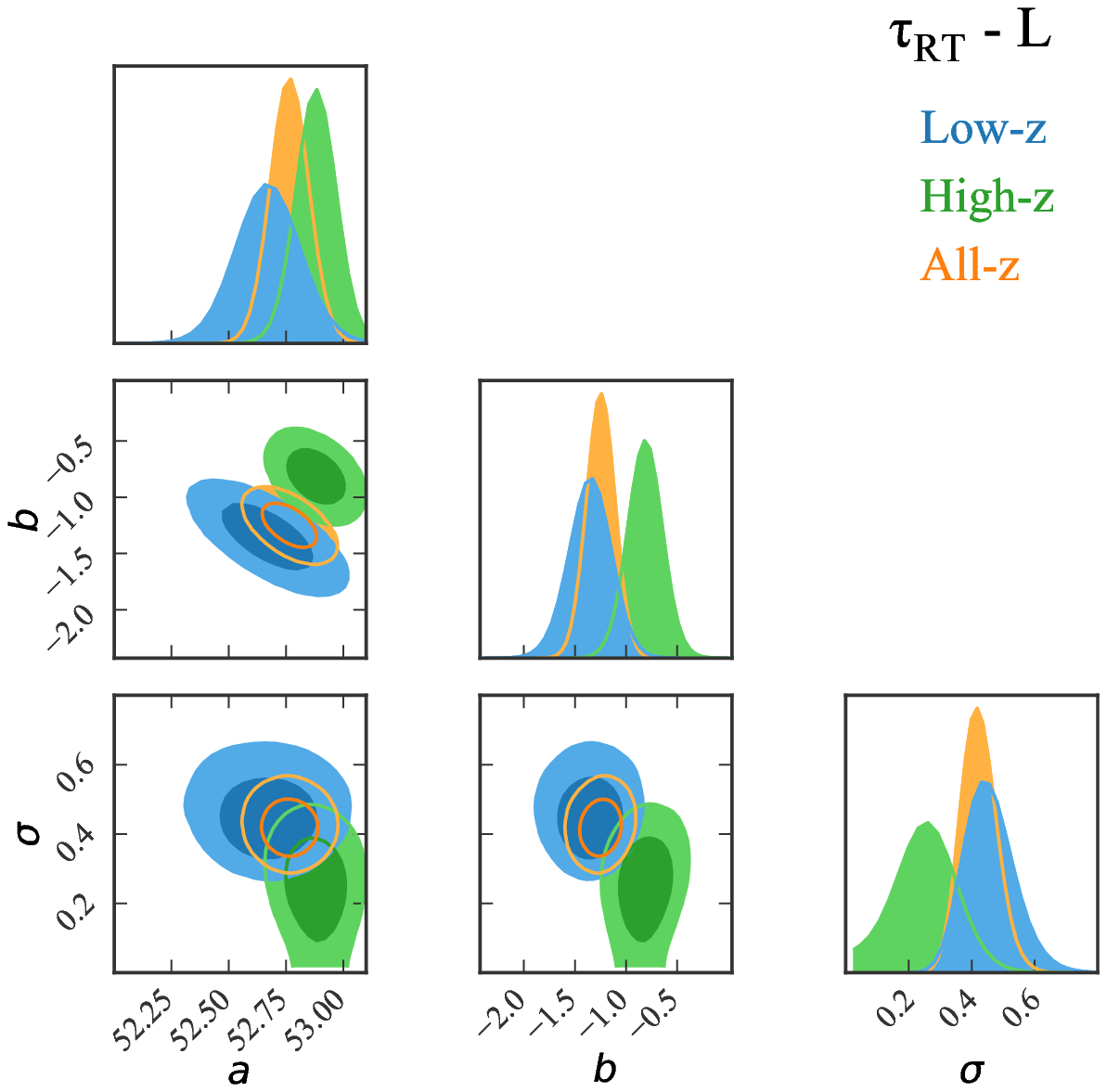}
\includegraphics[width=0.44\textwidth]{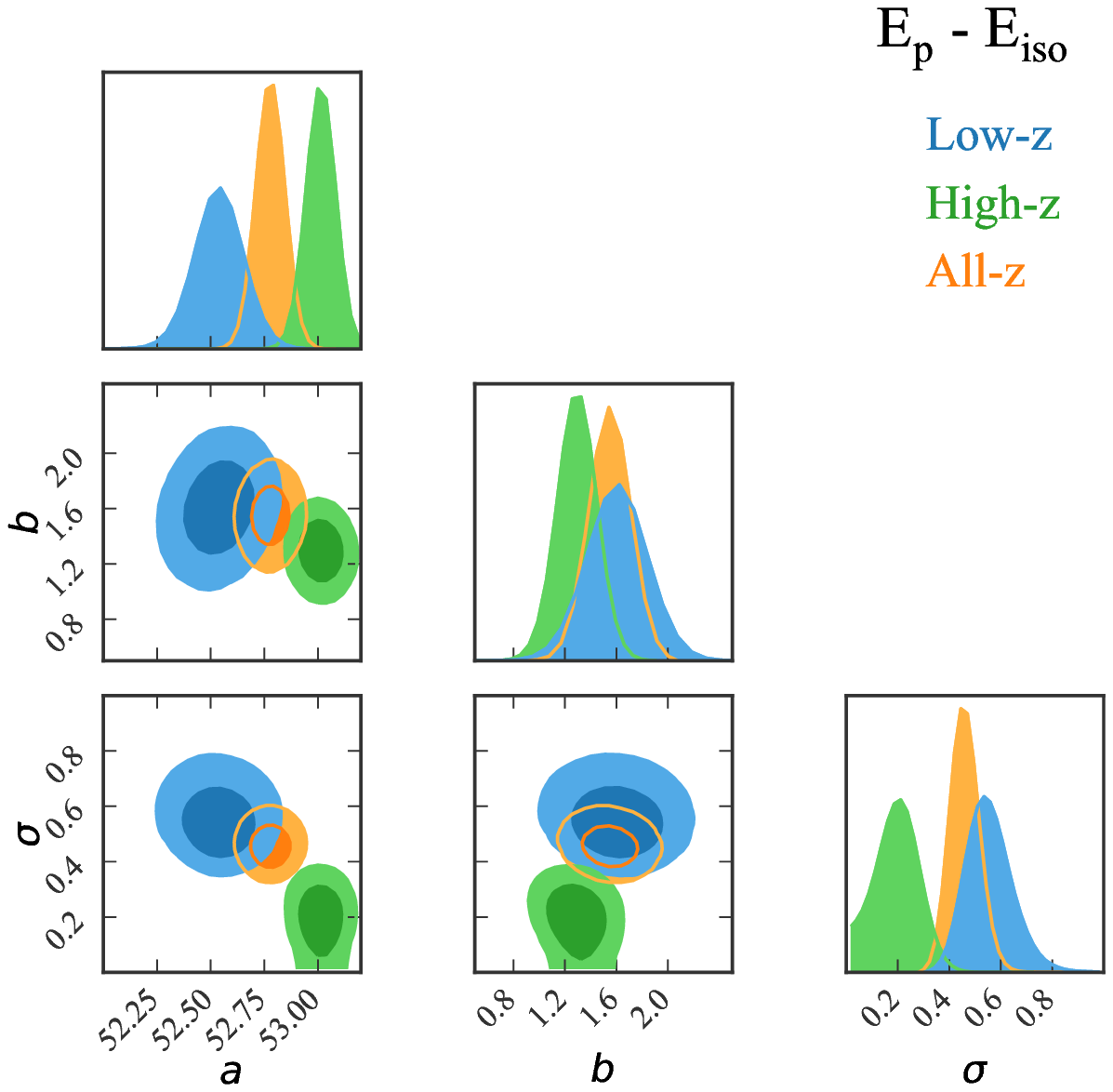}
\caption{\small{The confidence contours and marginalized PDFs for the parameters.}}\label{fig_contour}
\end{figure}

We perform a Markov Chain Monte Carlo analysis to calculate the posterior probability density function (PDF) of parameter space. We assume a flat prior on all the free parameters and limit $\sigma_{\rm int}>0$. Note that not all GRBs can be used to analyze each luminosity correlation, because not all the necessary quantities are measurable for some GRBs. For example, GRBs without measurement of the spectrum lag can not used in the $\tau_{\rm lag}-L$ analysis. Hence, we present the best-fitting parameters, together with the number of available GRBs in each fitting in Table \ref{tab:parameters}. In Figure \ref{fig_xy}, we plot all the six luminosity correlations in logarithmic coordinates. Low-$z$ and high-$z$ GRBs are represented by blue and red dots with the error bars denoting 1$\sigma$ uncertainties. The blue line, red line and black line stand for the best-fitting results for low-$z$ GRBs, high-$z$ GRBs and all-$z$ GRBs, respectively. The 1$\sigma$ and 2$\sigma$ contours and the PDFs for parameter space are plotted in Figure \ref{fig_contour}.

As shown in Table \ref{tab:parameters}, low-$z$ GRBs have a smaller intercept, but a sharper slope than high-$z$ GRBs for all the six luminosity correlations. All-$z$ GRBs have the parameter values between that of low-$z$ and high-$z$ subsamples. For the intrinsic scatter, low-$z$ GRBs have larger value than high-$z$ GRBs, and the $E_p-E_{\gamma}$ relation has the smallest intrinsic scatter hence we can only obtain its upper limit. The $V-L$ relation has the largest intrinsic scatter, thus it can not be fitted well with a simple line, which is legible in Figure \ref{fig_xy}. In Figure \ref{fig_contour}, the contours in the ($a,b$) plane indicate that the $E_p-E_{\gamma}$ relation of low-$z$ GRBs is consistent with that of high-$z$ GRBs at 1$\sigma$ confidence level. For the rest luminosity correlations, however, the intercepts and slopes for low-$z$ GRBs differ from that of high-$z$ GRBs at more than 2$\sigma$ confidence level.

\begin{figure}[htbp]
  \centering
  \includegraphics[width=0.8\textwidth]{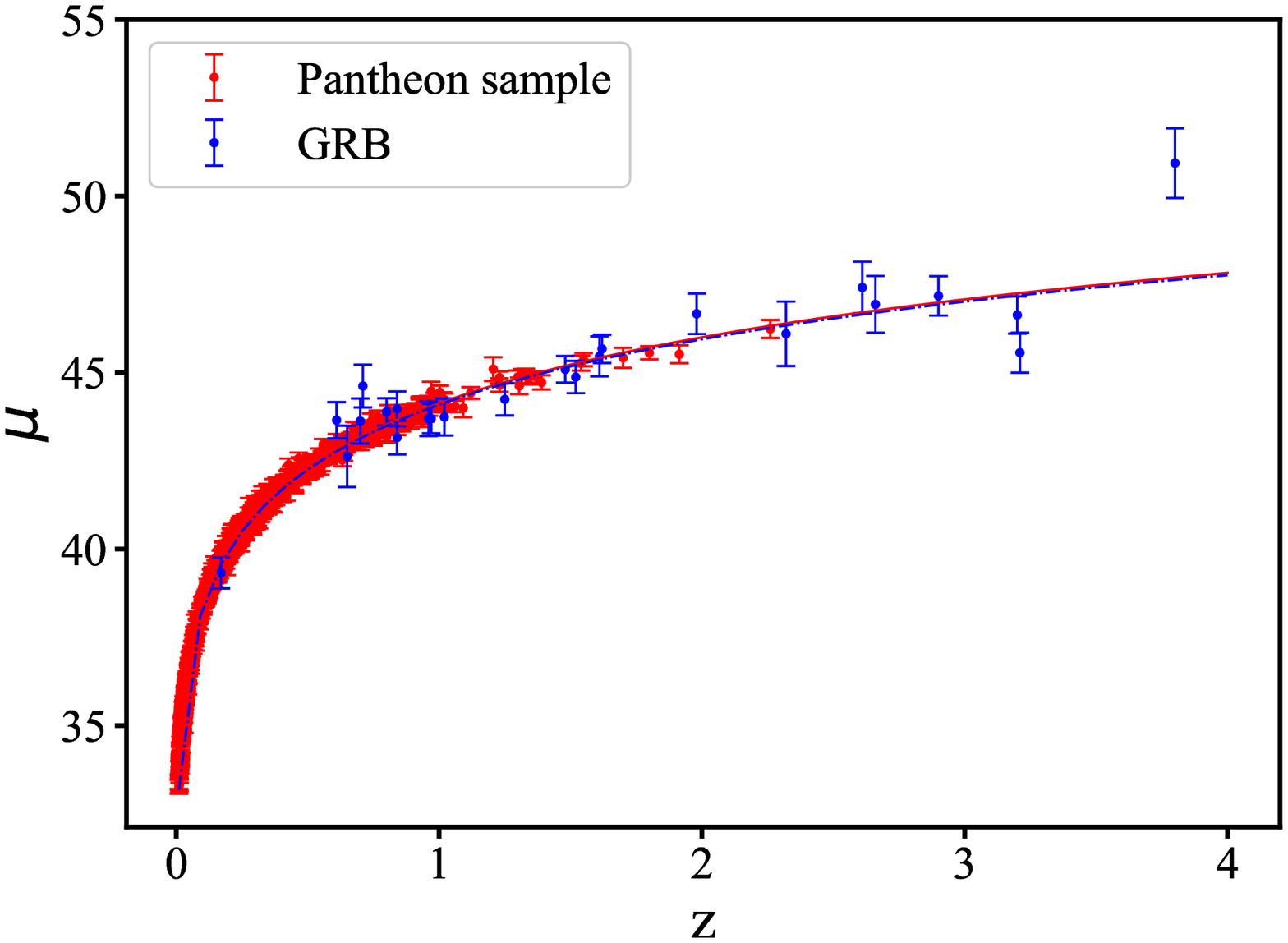}
  \caption{\small{The Hubble diagram of GRBs. The red dots with 1$\sigma$ error bars are the Pantheon data points. The blue dots with with 1$\sigma$ error bars are 24 long GRBs calibrated using the $E_p-E_{\gamma}$ relation. The lines are the best-fitting results to $\Lambda$CDM model with Pantheon (red line) and GRBs (blue dot-dashed line), respectively. Note: the two best-fitting lines seem to overlap with each other.}}\label{fig_fitLCDM}
\end{figure}

Having luminosity correlations calibrated, we can conversely using these correlations to calibrate the distance of GRBs, and further use GRBs to constrain cosmological models. Since our calibration of luminosity correlations is independent of cosmological model, the circularity problem is avoided. As we have seen, the $E_p-E_{\gamma}$ relation is not significantly evolving with redshift, so we use this relation to calibrate the distance of GRBs. Due to that the parameters of $E_p-E_{\gamma}$ relation in three samples (low-$z$, high-$z$ and full-$z$ samples) are consistent with each other, we directly apply the parameters obtained with all-$z$ sample. The distance of 24 GRBs calibrated using $E_p-E_{\gamma}$ are shown in the Hubble diagram in Figure \ref{fig_fitLCDM}. With the Pantheon dataset, the matter density of the flat $\Lambda$CDM model is constrained to be $\Omega_{\rm M}$=0.278$^{+0.007}_{-0.008}$. With 24 long GRBs alone, the matter density is constrained to be $\Omega_{\rm M}$=0.307$^{+0.065}_{-0.073}$. It indicates that the Hubble diagram in high redshift is consistent with the $\Lambda$CDM model.

\section{Discussions and conclusion}\label{sec:conclusion}

We have investigated the redshift dependence of six luminosity correlations in long GRBs using deep learning. We first reconstructed the distance-redshift relation to high redshift from the Pantheon compilation using RNN, and derived the uncertainty using BNN. Then the luminosity distance of GRBs is obtained without any assumption about the cosmological model. To test the possible redshift dependence of luminosity correlations, we divided GRBs into low-$z$ and high-$z$ subsamples and investigate the correlations for each subsample separately. It is found that, for all six luminosity correlations, low-$z$ subsample has a smaller intercept, but a sharper slope than high-$z$ subsample. In four out of six correlations ($\tau_{\rm lag} - L$, $E_p - L$, $\tau_{\rm RT} - L$ and $E_p - E_{\rm iso}$), the intercept and slope for low-$z$ subsample and high-$z$ subsample differ at more than $2\sigma$. For the $V-L$ relation, the intrinsic scatter is too large to make a reliable conclusion. For the $E_p-E_{\gamma}$ relation, there is no evidence for redshift evolution, and the intrinsic scatter is the smallest among the six correlations. However, the number of available GRBs for the $E_p-E_{\gamma}$ relation is small, because most long GRBs lack the measurement of jet opening angle which is necessary to calculate the beaming factor. The constraint on flat $\Lambda$CDM model from 24 GRBs calibrated using $E_p-E_{\gamma}$ relation gives $\Omega_{\rm M}$=0.307$^{+0.065}_{-0.073}$, which is well consistent with the result constrained from Pantheon. Our network is trained using Pantheon sample, so GRBs calibrated using the network should be consistent with Pantheon sample.

In the RNN+BNN network, the network is trained using the Pantheon data to extend the Hubble diagram to high redshift range without any assumption about the cosmological model. In the deep learning method, the kernel is achieving the most efficient expression to denote the relationship between the features and the targets of Pantheon data, where the validity is reflected in the architecture of network. In the cooperation of the Monte Carlo dropout and the activation function, our network not only predicts the central value, but also the corresponding confidence region. Because the number of SNe Ia whose redshift beyond 1.4 is small, the uncertainty in the poor observational data range would be large. Therefore, we choose a small dropout rate 0.2 with the specific activation function $A_{f_{\rm Tanh}}$ to acquire a relatively small uncertainty.

The method proposed here has some advantages compared with other methods already existing in literatures. \citet{Lin:2016} have tested the redshift-dependence of six luminosity correlations by dividing GRBs into low-$z$ and high-$z$ subsamples. In their work, the slope and intercept parameters $(a,b)$ are calculated based on a specific cosmological model. In contrast, our method proposed is completely independent on cosmological model. Several works have been devoted to calibrating the distance of GRBs in a model-independent way \citep{Firmani:2005,Liang:2005,Capozziello:2010,Wei:2010,Liu:2015}. All of these methods rely on the assumption that the luminosity correlations have no redshift evolution, which couldn't be tested by the method itself. To calibrate high-$z$ GRBs, one must first obtain the luminosity correlation coefficients $(a,b)$ from the low-$z$ GRB, then directly extrapolate the correlation to high-$z$ GRBs. \citet{{Wang:2011}} proposed to calibrate GRBs by fitting the coefficients $(a,b)$ and the cosmological parameters simultaneously to a specific cosmological model, so the circularity problem can be avoided. This calibrating method is still model-dependent, and the GRB calibrated in one cosmological model couldn't be used to constrain other models. To constrain other cosmological models, the coefficients $(a,b)$ should be refitted again. In comparison, our method can be used to test the redshift evolution of luminosity correlations model-independently. If a correlation is redshift-independent, low-$z$ and high-$z$ GRBs can be calibrated simultaneously using this correlation. GRBs calibrated in this way can be directly used to constrain cosmological models.

\begin{acknowledgments}
This work has been supported by the National Natural Science Fund of China Grant Nos. 11603005, 11775038 and 11947406, and the Fundamental Research Funds for the Central Universities Grant No. 2020CQJQY-Z003.
\end{acknowledgments}

\end{document}